 \documentclass[10pt,preprint]{aastex}  
 \usepackage{natbib}
\def\ang{\AA}
\def\arcsec{\hbox{$^{\prime\prime}$}}

\def\gapprox{\lower.4ex\hbox{$\;\buildrel >\over{\scriptstyle\sim}\;$}}
\def\lapprox{\lower.4ex\hbox{$\;\buildrel <\over{\scriptstyle\sim}\;$}}

\shortauthors{ASCHWANDEN ET AL. 2012}
\shorttitle{STEREOSCOPY OF CORONAL LOOPS: IV}

\begin{document}

\title{		First 3D Reconstructions of Coronal Loops with the STEREO 
		A+B Spacecraft: IV. Magnetic Modeling with Twisted
		Force-Free Fields
		\footnote{Manuscript version, 2012-Jul-4}}

\author{        Markus J. Aschwanden, Jean-Pierre Wuelser, Nariaki V. Nitta,
		James R. Lemen, Marc L. DeRosa, and Anna Malanushenko }

\affil{         Lockheed Martin Advanced Technology Center,
                Solar \& Astrophysics Laboratory,
                Org. ADBS, Bldg.252,
                3251 Hanover St.,
                Palo Alto, CA 94304, USA;
                e-mail: aschwanden@lmsal.com}

\begin{abstract}
The three-dimensional (3D) coordinates of stereoscopically triangulated
loops provide strong constraints for magnetic field models of
active regions in the solar corona. Here we use STEREO/A and B data from
some 500 stereoscopically triangulated loops observed in four active
regions (2007 Apr 30, May 9, May 19, Dec 11), together with SOHO/MDI 
line-of-sight magnetograms. We measure the average misalignment angle
between the stereoscopic loops and theoretical magnetic field models,
finding a mismatch of $\mu=19^\circ-46^\circ$ for a potential field model,
which is reduced to $\mu=14^\circ-19^\circ$ for a non-potential field model 
parameterized by twist parameters.
The residual error is commensurable with stereoscopic measurement
errors ($\mu_{SE} \approx 8^\circ-12^\circ$). We developed a potential
field code that deconvolves a line-of-sight magnetogram into three
magnetic field components $(B_x, B_y, B_z)$, as well as a non-potential
field forward-fitting code that determines the full 
length of twisted loops ($L \approx 50-300$ Mm),
the number of twist turns (median $N_{twist}=0.06$),
the nonlinear force-free $\alpha$-parameter 
(median $\alpha \approx 4 \times 10^{-11}$ cm$^{-1}$),
and the current density (median $j_z \approx 1500$ Mx cm$^{-2}$ s$^{-1}$).
All twisted loops are found to be far below the critical value
for kink instability, and Joule dissipation of their currents 
is found be be far below the coronal heating requirement.
The algorithm developed here, based on an analytical solution of
nonlinear force-free fields that is accurate to second order 
(in the force-free parameter $\alpha$),
represents the first code that enables fast forward-fitting 
to photospheric magnetograms and stereoscopically triangulated loops
in the solar corona.
\end{abstract}


\section{          	    INTRODUCTION  			}

This is paper IV of a series that explores the three-dimensional (3D)
reconstruction of coronal loops with data from the two STEREO A and B
spacecraft. The previous three studies focused on the 3D geometry of coronal
loops (Aschwanden et al.~2008b), on electron density and temperature
measurements (Aschwanden et al.~2008c), and on active region modeling
with a stereoscopic tomography method (Aschwanden et al.~2009). In this
fourth study we concentrate on magnetic modeling of four stereoscopically
observed active regions, which should reveal whether potential or
non-potential magnetic field models fit the observed data better, and
should provide more accurate information on magnetic fields and
electric currents that are directly inferred from observables, and this way
allow us to test theoretical magnetic field models of the solar corona.
The crucial benefit of this study is the inference of the true 3D magnetic
field geometry, which presently can only be accurately measured by
stereoscopic triangulation with the dual spacecraft STEREO/A and B,
and thus it provides a most crucial input to test theoretical magnetic
field models.

The {\sl Solar TErrestrial RElations Observatory (STEREO)} was launched
on October 26, 2006, consisting of two identical spacecraft A(head) and
B(ehind) that orbit from the Earth in opposite directions around the Sun,
with a separation
angle that increases by $\approx 45^\circ$ per year (Kaiser 2008).
Some tutorials or reviews on magnetic modeling using STEREO data can be
found in Inhester (2006), Wiegelmann et al.~(2009), and Aschwanden (2011).
Early methods of stereoscopic 3D magnetic field modeling have been already
proposed in the pre-STEREO era, including loop twist measurements by fitting
the projected shape of curved helical field lines (Portier-Fozzani
et al.~2001), by fitting a linear force-free model to coronal loops
geometrically reconstructed with solar-rotation stereoscopy (Wiegelmann
and Neukirch 2002; Feng et al.~2007a), by tomographic reconstruction
constrained by the magnetohydrostatic equations (Wiegelmann and Inhester 2003;
Ruan et al.~2008),
by comparison of theoretical magnetic field models with spectro-polarimetric
loop detections (Wiegelmann et al.~2005), or by using theoretical magnetic 
field models to resolve the stereoscopic correspondence ambiguity of 
triangulated loops (Wiegelmann and Inhester 2006; Conlon and Gallagher 2010). 
Once the STEREO mission was launched, the first stereoscopic triangulations 
of loops were performed from data of April 2007 onward, with a spacecraft 
separation angle of $\alpha_{sep} \gapprox 5^{\circ}$ (Feng et al.~2007b;
Aschwanden et al.~2008b). First fits of linear force-free field
extrapolations (using SOHO/MDI magnetograms) to stereoscopically 
triangulated loops were found to be offset (Feng et al.~2007b;
Inhester et al.~2008). A benchmark test of one potential field code and
11 nonlinear force-free field codes modeling a partial active region 
(NOAA 10953) observed with Hinode/SOT and STEREO/EUVI on 2007 Apr 30 revealed
mean 3D misalignment angles of $\mu=24^\circ$ for the potential
field code, and a range of $\mu=24^{\circ}-44^\circ$ for the 
non-potential field (NLFFF) codes, which was attributed to the
non-force-freeness in chromospheric heights and uncertainties in the
boundary data (DeRosa et al.~2009). Misalignment measurements between
stereoscopically triangulated loops and {\sl potential field source
surface (PFSS)} magnetic field extrapolations were then extended to
four active regions with a similar finding of $\mu = 19^\circ-36^\circ$
(Sandman et al.~2009). The misalignment could be reduced by about a factor
of two for potential field models that were not constrained by observed
magnetograms, where the magnetic field is parameterized by buried
unipolar magnetic charges (Aschwanden and Sandman 2010) or by 
a small number of 3-5 submerged dipoles (Sandman and Aschwanden 2011).
These exercises demonstrated that potential field solutions exist that
match the 3D geometry of coronal loops better than standard extrapolations
from photospheric magnetograms. 

Modeling of non-potential magnetic fields, constrained by stereoscopically 
triangulated loops, however, is still unexplored. A first step in this
direction was pioneered by Malanushenko et al.~(2009), taking a particular
model of a nonlinear force-free field with a known analytical solution 
(Low and Lou 1990) and fitting a linear force-free field individually to
each field line, which yields the twist and force-free $\alpha$-parameter 
for each loop individually, but not in form of a self-consistent nonlinear
force-free field. In a next step, a self-consistent nonlinear force-free
field was fitted to the $\alpha$'s or each loop (Malanushenko et al.~2012).
In this study we pursue for the first time non-potential magnetic field 
modeling applied to an observed set of coronal loops, i.e., to some 500 loops 
that have been stereoscopically triangulated from four different active regions.
Since currently available NLFFF codes are very computing-intensive, 
which precludes an iterative fitting to individual loops, we developed a
magnetic field parameterization that is suitable for fast forward-fitting. 
In this parameterization, the magnetic field is characterized by a 
superposition of point charges with vertically twisted fields, which is 
approximately force-free (to second order in $\alpha$), as derived
analytically (Aschwanden 2012) and tested numerically (Aschwanden and 
Malanushenko 2012). Since forward-fitting requires some form of model 
parameterization, tests with physics-based models (such as twisted field 
components used here) will be useful to explore how closely the observed 
loop geometry can be fitted at all, although our choice of parameterization
represents only a subset of all possible nonlinear force-free solutions.
We will test the validity of the divergence-freeness and force-freeness 
numerically and quantify it by common figures of merit, which can be 
compared with the performance of other full-fledged NLFFF codes (which,
however, are not capable of fast forward-fitting to observations).  

The outline of this paper is as follows: Section 2 contains the 
theoretical outline of a force-free field approximation, Section 3 presents
the results of forward-fitting to some 500 stereoscopically triangulated
loops, and Section 4 concludes with a discussion of the results. 
Technical details of the forward-fitting code are described in 
Appendix A and in two related documentations (Aschwanden 2012; 
Aschwanden and Malanushenko 2012).

\section{         THEORETICAL MODEL AND DATA ANALYSIS METHOD 	}

Here we develop a novel method to determine an approximate 3D magnetic
field model of a solar active region, using a magnetogram that contains
the line-of-sight magnetic field component and a set of stereoscopically 
triangulated loops, observed with STEREO/A and B in extreme ultraviolet 
(EUV) wavelengths. A flow chart of this magnetic field modeling 
algorithm is given in Fig.~1. 

\subsection{	3D Potential Field with Unipolar Magnetic Charges 	}

Magnetograms that contain the line-of-sight magnetic field component
$B_z(x,y)$ are readily available, while vector magnetograph data that
provide the 3D magnetic field components $[B_x(x,y), B_y(x,y), B_z(x,y)]$
are only rarely available, are difficult to calibrate, resolving the 
$180^\circ$ ambiguity is challenging (even with the recent HMI/SDO 
instrument), and they generally contain substantially larger data noise 
in the B-components than magnetograms. 
It is therefore desirable to develop a method that
infers 3D magnetic field components from magnetogram images. Such a 
method was recently developed with the concept of parameterizing the
3D magnetic field with subphotospheric unipolar magnetic charges that
can be determined from an observed line-of-sight magnetogram, as described
in Aschwanden and Sandman (2010). In the first version we neglected the
curvature of the solar surface, because the analyzed active regions have
generally a substantially smaller size than the solar radius. For higher
accuracy of the magnetic field model, however, we generalize the method
by including here the full 3D geometry of the curved solar surface.

The simplest representation of a magnetic potential field 
that fulfills Maxwell's divergence-free condition ($\nabla \cdot {\bf B}=0$) 
is a magnetic charge that is buried below the solar surface (to avoid
magnetic monopoles in the corona), which predicts a magnetic field
${\bf B}({\bf x})$ that points away from the buried unipolar charge
and whose field strength falls off with the square of the distance $r$,
\begin{equation}
        {\bf B}({\bf x})  
        = B_0 \left({d_0 \over r}\right)^2 {{\bf r} \over r} \ ,
\end{equation}
where $B_0$ is the magnetic field strength at the solar surface directly
above the buried magnetic charge, ${\bf r}_0=(x_0,y_0,z_0)$ is the
subphotospheric position of the buried charge, $d_0=\sqrt{1-x_0^2-y_0^2-z_0^2}$ 
is the depth of the magnetic charge, and ${\bf r}=[(x-x_0), (y-y_0), (z-z_0)]$ 
is the distance vector of an arbitrary location ${\bf x}=(x,y,z)$ in the solar 
corona (where we desire to calculate the magnetic field) from the location 
${\bf r}_0$ of the buried charge. We choose a cartesian coordinate system 
$(x,y,z)$ with the origin in the Sun center and use units of solar 
radii, with the direction of $z$ chosen along the line-of-sight from Earth
to Sun center.  For a location near disk center ($x \ll 1, y \ll 1$), the 
magnetic charge depth is $d_0 \approx (1-z_0)$, an approximation that was used 
earlier (Aschwanden and Sandman 2010), while we use here the exact 3D distances
to account for the curvature of the solar surface and for off-center 
positions of active region loops. 

Our strategy is to represent an arbitrary line-of-sight magnetogram with a
superposition of $N_m$ magnetic charges, so that the potential field can be
represented by the superposition of $N_m$ fields 
${\bf B}_j$ from each magnetic charge $j=1,...,N_m$,
\begin{equation}
        {\bf B}({\bf x}) = \sum_{j=1}^{N_m} {\bf B}_j({\bf x})
        = \sum_{j=1}^{N_m}  B_j
        \left({d_j \over r}\right)^2 {{\bf r} \over r} \ ,
\end{equation}
with ${\bf r}=[(x-x_j), (y-y_j), (z-z_j)]$ 
the distance from the magnetic charge $j$.
Since the divergence operator is linear, the superposition of a number of
potential fields is divergence-free also, 
\begin{equation}
	\nabla \cdot {\bf B} = \nabla \cdot (\sum_j {\bf B}_j) 
	= \sum_j (\nabla \cdot {\bf B}_j) = 0 \ . 
\end{equation}
This way we can parameterize a 3D magnetic field ${\bf B}({\bf x})$ (Eq.~2)
with $4 \times N_m$ parameters, i.e., $(x_j, y_j, z_j, B_j), j=1,...,N_m$. 

A simplified algorithm to derive the magnetic field parameters 
$(x_j, y_j, z_j, B_j), j=1,...,N_m$ from a line-of-sight magnetogram
$B_z(x,y)$ is described in the previous study (Aschwanden and Sandman 2010). 
Essentially we start at the position $(x_1, y_1)$ of the peak magnetic 
field $B_z(x_1, y_1)$ in the magnetogram image, fit the local magnetic field 
profile ${\bf B}_1({\bf x})$ (Eq.~1) to obtain the parameters of the 
first magnetic field component $(x_1, y_1, z_1, B_1)$, subtract the first
component from the image $B_z(x,y)$, and then iterate the same procedure
at the second-highest peak to obtain the second component 
$(x_2, y_2, z_2, B_2)$, and so forth, until we stop at the last component 
$(x_{N_m}, y_{N_m}, z_{N_m}, B_{N_m})$ above some noise threshold level.
We alternate the magnetic polarity in the sequence of subtracted magnetic
charge components, in order to minimize the absolute value of the
residual maps. 
The number of necessary components typically amounts to $N_m \approx 100$ 
for an active region, depending on the complexity and size of the active
region. The geometrical details of the inversion of the parameters 
$(x_j, y_j, z_j, B_j)$ of a magnetic field component from the observables
$(B_z, x_p, y_p, w)$ of an observed peak in a line-of-sight magnetogram
at position $(x_p, y_p)$ with peak value $B_z$ and width $w$ is
derived in Appendix A. 

\subsection{	Force-free Magnetic Field Model   		   }

With the 3D parameterization of the magnetic field described in the foregoing
section we can compute magnetic potential fields that fulfill Maxwell's
divergence-free condition and are ``current-free'',
\begin{equation}
		\begin{array}{rl}
		\nabla \cdot {\bf B} &= 0 \\
		{\bf j}/c &= {1 \over 4\pi} (\nabla \times {\bf B}) = 0 \ . \\
		\end{array} 
\end{equation}
However, magnetic potential field models do not fit observed 3D loop
geometries sufficiently well (DeRosa et al.~2009; Sandman et al.~2009;
Aschwanden and Sandman 2010; Sandman and Aschwanden 2011).
Therefore, we turn now to non-potential magnetic field models of the
type of nonlinear force-free fields (NLFFF), 
\begin{equation}
		\begin{array}{rl}
		\nabla \cdot {\bf B} &= 0 \\
		{\bf j}/c &= {1 \over 4\pi} (\nabla \times {\bf B}) = 
		\alpha({\bf x}) {\bf B} \\
		\end{array} \ ,
\end{equation}
where $\alpha({\bf x})$ is a scalar function that varies in space,
but is constant along a given field line, and the current ${\bf j}$ 
is co-aligned and proportional to the magnetic field ${\bf B}$. 
General NLFFF solutions of Eq.~(5), however, are very computing-intensive
and are subject to substantial uncertainties due to insufficient boundary
constraints and non-force-free conditions in the lower chromosphere
(Schrijver et al.~2006; DeRosa et al.~2009). In our approach to model
NLFFF solutions that fit stereoscopically triangulated loops we thus choose
approximate analytical NLFFF solutions that can be forward-fitted
to the observed 3D loop geometries much faster. The theoretical formulation
of the force-free magnetic field approximation suitable for fast forward-fitting
is described in Aschwanden (2012) and numerical tests are given in
Aschwanden and Malanushenko (2012).

The essential idea is that we generalize the concept of buried point charges
we used for the representation of a potential field, by adding an azimuthal 
magnetic field component $B_{\varphi}(\rho)$ that describes a twist around a 
vertical axis. The concept is visualized in Fig.~2. An untwisted flux tube
can be represented by parallel field lines $B_s(\rho)$ along an axis $s$
with variable distances $\rho$ from a given axis $s$ (Fig.~2, left).
A uniformly twisted flux tube with some azimuthal component 
$B_{\varphi}(\rho)$ around the twist axis $s$ (Fig.~2, middle) has the 
following analytical solution for a force-free field (Gold and Hoyle 1958; 
Priest 1982, p.125; Sturrock 1994, p.216;
Boyd and Sanderson 2003, p.102; Aschwanden 2004, p.216), 
\begin{equation}
        B_s(\rho) = {B_0 \over 1 + b^2 \rho^2} \ ,
\end{equation}
\begin{equation}
        B_{\varphi}(\rho) = {B_0 b \rho \over 1 + b^2 \rho^2} \ ,
\end{equation}
where $b$ is a parameter that quantifies the number $N_{twist}$ of full twist 
turns over a (loop) length $L$,
\begin{equation}
	b = {2 \pi N_{twist} \over L} , 
\end{equation}
and is related to the force-free parameter $\alpha$ in the force-freeness condition
$(\nabla \times {\bf B}) = \alpha({\bf x}) {\bf B}$ by, 
\begin{equation}
        \alpha (\rho) = {2 b \over (1 + b^2 \rho^2)} \ .
\end{equation}
Thus, we see that $\alpha$ has a finite value $\alpha(\rho=0)=2b$ at the twist
axis and drops monotonically with increasing distance $\rho$ from the twist axis.

In a next step we generalize the concept of a cylindrical twisted flux tube
(Fig.~2, middle) to a point charge with an associated vertically twisted field 
(Fig.~2, right). The main difference to the twisted flux tube, which has a
constant cross-section and thus a constant magnetic flux along the cylinder axis,
is the quadratically decreasing field strength of the longitudinal field component
$B_s(s)$ along the twist axis with distance $s$, to conserve the magnetic flux.
It can be shown analytically, that this topology has the following approximate
force-free and divergence-free solution (satisfying Eqs.~4 and 5), if we neglect
second-order and higher-order terms of $[b r \sin(\theta)]$ (Aschwanden 2012),
\begin{equation}
        B_r(r, \theta) = B_0 \left({d^2 \over r^2}\right)
        {1 \over (1 + b^2 r^2 \sin^2{\theta})} \ ,
\end{equation}
\begin{equation}
        B_\varphi(r, \theta) =
        B_0 \left({d^2 \over r^2}\right)
        {b r \sin{\theta} \over (1 + b^2 r^2 \sin^2{\theta})} \ ,
\end{equation}
\begin{equation}
        B_\theta(r, \theta) \approx 0
        \ ,
\end{equation}
\begin{equation}
        \alpha(r, \theta) \approx {2 b \cos{\theta} \over
        (1 + b^2 r^2 \sin^2{\theta})}  \ .
\end{equation}
Since $b \propto \alpha$ for $\theta \ll 1$ (Eq.~13), we refer to this 
approximation as second-order accuracy in $\alpha$ for short.
Thus we can characterize a point charge with a vertically twisted field with
5 parameters: $(B_j, x_j, y_j, z_j, \alpha_j)$, where the force-free parameter 
$\alpha_j = \alpha(\rho=0)$ at the twist axis is related to $b_j$ by 
$\alpha_j = 2 b_j$. 

In analogy to the superposition principle we used to construct an arbitrary
potential field based on a number of $N_m$ magnetic charges, we apply the
same superposition rule to a number of $N_m$ point charges with vertical
twist,
\begin{equation}
        {\bf B}({\bf x}) = \sum_{j=1}^{N_m} {\bf B}_j({\bf x}) \ ,
\end{equation}
and we can parameterize an arbitrarily complex non-potential field with 
$5 \times N_m$ free parameters. The superimposed field ${\bf B}$, which
consists of a linear combination of quasi-force-free field components
(to second order in $\alpha$) is not exactly force-free, but it
can be shown that the NLFFF approximation expressed in Eq.~(10-13) is 
divergence-free to second-order accuracy in the term $[b r \sin(\theta)]$,
and force-free to third-order accuracy (Aschwanden 2012), which we call 
``quasi-force-free'' here.
The superposition of multiple twisted sources, which have each a twist axis
with a different location and orientation (due to the curvature of the solar
surface), needs to be transformed into a common cartesian coordinate
system, which is derived in detail in Section 2.4 in Aschwanden (2012). 

We show a few examples of this non-potential magnetic field model in Fig.~3,
for a unipolar, a dipolar, and a quadrupolar configuration. More cases are
simulated in Aschwanden (2012). Three cases are shown for the
non-potential case with twist (Fig.~3, right panels), which degenerate
to the potential field case if the twist parameter is set to zero 
($\alpha_j=0$) (Fig.~3, left panels).

In our data analysis we perform a global fit of our non-potential field model
to a set of stereoscopically triangulated loops, Typically we use a magnetic
field model with $N_m \le 200$ magnetic field components (which has
$\le 800$ potential field parameters that can directly be inverted from the
line-of-sight magnetogram $B_z(x,y,z_{ph})$, and $\le 200$ non-potential
field parameters $\alpha_j$, which need to be forward-fitted to the data,
consisting of the stereoscopic 3D coordinates of some $N_L \approx 70-200$ 
loops per active region.
Each forward-fit minimizes the average misalignment angle (along and among the
loops) between the theoretical model (i.e., twisted fields) 
and the observed loop (with stereoscopically triangulated 3D coordinates).
The misalignment angle at a position ${\bf x}$ is defined as
(Sandman et al.~2009; Aschwanden and Sandman 2010; Sandman and 
Aschwanden 2011):
\begin{equation}
        \mu({\bf x}) =
        cos^{-1} \left({ {\bf B}^{theo}({\bf x}) \cdot
        {\bf B}^{obs}({\bf x}) \over
        |{\bf B}^{theo}({\bf x})|\ |{\bf B}^{obs}({\bf x})| }\right) \ .
\end{equation}
We derive a characteristic misalignment angle from measurements at 10 
equi-spaced postions ${\bf x}$ along each observed loop segment, 
where a unique theoretical field line ${\bf B}_{theo}({\bf x})$ is
calculated at each loop position. Generally, this yields 10 theoretical
field lines that intersect a single loop at the 10 chosen positions,
which degenerate to one single theoretical field line in the case of
a perfectly matching model. The range of misalignment angles ignores the 
$180^\circ$-ambiguity and is defined in the range of 
$\mu=[0^\circ, 90^\circ]$.
Technical details of the numerical algorithm of our forward-fitting code 
to the stereoscopic loop coordinates are provided in Aschwanden and
Malanushenko (2012). 

\subsection{	Stereoscopical Triangulation of Loops			}

In order to obtain the 3D coordinates of coronal loops we require one (or 
multiple wavelength filter) image pairs recorded with the two spacecraft
STEREO/A(head) and B(ehind), at some spacedraft separation angle that
is suitable for stereoscopy. Here we will use STEREO data from the
first year (2007) of the mission, when the spacecraft separation was
$\alpha_{sep} \lapprox 45^\circ$. An example of such a stereoscopic
pair of EUVI images recorded at a wavelength of 171 \ang\ is shown in
Fig.~1 (top right). Loop recognition requires a segmentation algorithm
that traces curvi-linear features. There are visual/manual methods as well as
automated methods, which have been compared in a benchmark study
(Aschwanden et al.~2008a). Although automated detection algorithms have
improved to the level or approaching visual perception for single images
(Aschwanden 2010), an automated algorithm for dual detection of 
corresponding loop patterns in 
stereoscopic image pairs has not yet been developed yet, so that we have
to rely on manual/visual detection. The 3D coordinates of such visually
detected loops have been documented in previous publications
(Aschwanden 2008b,c, 2009; Sandman et al.~2009; Aschwanden and Sandman 
2010; Sandman and Aschwanden 2011).

The technique of loop detection falls into the category of image segmentation
algorithms. The easiest way to segment coronal loops in an EUV image is
to apply first a highpass filter (Fig.~1), which enhances fine structures,
in particular coronal loops with narrow widths. Stereoscopic triangulation
can then be accomplished by (1) co-aligning a stereoscopic image pair into
epipolar coordinates (where the stereoscopic parallax effect has the
same direction in both images), (2) tracing a loop visually in a STEREO/A 
image, (3) identifying the corresponding counterpart in the STEREO/B image
using the expected projected altitude range;
(4) stereoscopic triangulation of the loop coordinates 
$(x_L[s],y_L[s],z_L[s])$ (along the loop length coordinate $s$),
and (5) coordinate transformations into an Earth-Sun heliographic coordinate
system. Here we have to transform STEREO/A coordinates into the SoHO/MDI
coordinate system, in order to enable magnetic modeling with MDI magnetograms.
The height range of stereoscopically triangulated loops does generally
not exceed 0.15 solar radii, due to the decrease of dynamic range in flux
for altitudes in excess of one hydrostatic scale height.
Here we used the loop tracings and stereoscopic measurements
from three different EUVI channels combined (171, 195, 284 \ang ).
The method of stereoscopic triangulation is described in detail in
previous papers (Aschwanden 2008b,c, 2009).  

What is also important to know when we attempt magnetic modeling of
stereoscopically triangulated coronal loops is the related uncertainty.
The parallax effect occurs mostly in east-west direction, which implies
that loop segments with North-South direction can be most accurately
triangulated, while the accuracy decreases progressively for segments
with longer components in east-west direction (Aschwanden 2008b). Another 
source of error is mis-identification or confusion of dual loop 
counterparts in stereoscopic image pairs, an uncertainty that can only 
be quantified in an empirical manner. Such empirical estimates of 
stereoscopic errors were evaluated by measuring deviations from
parallelity of loop pairs triangulated in near proximity, which yielded 
an estimated uncertainty in the absolute direction of $\mu_{SE}=8^\circ -
12^\circ$ for four analyzed active regions (Aschwanden and
Sandman 2010). So, even when we achieve a theoretical magnetic field
model that perfectly matches the observed loops, we expect a residual
misalignment error of $\mu \approx 10^\circ$ due to the
uncertainty of stereoscopic triangulation. 

\subsection{	3D Magnetic Field Representation 		}

The input from the line-of-sight magnetogram provides the longitudinal
component $B_z(x,y,z_{ph})$ at the photospheric level $r=1$, which 
corresponds to the z-coordinate $z_{ph}=\sqrt{(1-x^2-y^2)}$. 
The decomposition of the magnetogram data into unipolar magnetic
charges (Section 2.1 and Appendix A) yields the full 3D vector field,
${\bf B}^P({\bf x})=[B_x(x,y,z_{ph}), B_y(x,y,z_{ph}), B_z(x,y,z_{ph})]$ 
on the solar surface at the photospheric level $z_{ph}$, which is a potential
field by definition, since the derived field consists of a superposition
of potential fields from single magnetic charges (Eq.~2). 

The nonlinear force-free magnetic field model (Section 2.2), which we fit
to the data by optimizing the misalignment angles with our chosen
parameterization, defines a non-potential field 
${\bf B}({\bf x})=[B_x(x,y,z)$, $B_y(x,y,z)$, $B_z(x,y,z)]$, which can
be extrapolated at any point in the corona by stepping along a field line
$[x(s), y(s), z(s)]$ in small incremental steps $\Delta s$, where the
cosines of the cartesian magnetic field components 
$(B_x, B_y, B_z)$ define also the direction of the field line, so that 
it can be iteratively calculated with 
\begin{equation}
	\begin{array}{ll}
		x(s+\Delta s) &= x(s) + \Delta s [B_x(s)/B(s)] p \\
		y(s+\Delta s) &= y(s) + \Delta s [B_y(s)/B(s)] p \\
		z(s+\Delta s) &= z(s) + \Delta s [B_z(s)/B(s)] p \\
	\end{array} \ .
\end{equation}
where $p=\pm 1$ represents the sign or polarization of the magnetic charge.

The force-free parameter $\alpha(x, y, z)$ of our forward-fitted field
$B(x, y, z)$ at any given point of space in the computational box 
$(x_i, y_j, z_k)$ can be numerically computed for each of the three vector 
components of $(\nabla \times {\bf B})$,
\begin{equation}
        \alpha_x({\bf x}) = {1 \over 4 \pi}
        {(\nabla \times {\bf B})_x \over {\bf B}_x}
        = {1 \over 4 \pi B_x} \left({\partial B_z \over \partial y}
             - {\partial B_y \over \partial z}\right) \ ,
\end{equation}
\begin{equation}
        \alpha_y({\bf x}) = {1 \over 4 \pi}
        {(\nabla \times {\bf B})_y \over {\bf B}_y}
        = {1 \over 4 \pi B_y} \left({\partial B_x \over \partial z}
             - {\partial B_z \over \partial x}\right) \ ,
\end{equation}
\begin{equation}
        \alpha_z({\bf x}) = {1 \over 4 \pi}
        {(\nabla \times {\bf B})_z \over {\bf B}_z}
        = {1 \over 4 \pi B_z} \left({\partial B_y \over \partial x}
             - {\partial B_x \over \partial y}\right) \ .
\end{equation}
using a second-order scheme for the spatial derivatives,
i.e., $\partial B_x / \partial y = (B_{i+1j,k}-B_{i-1,j,k})
/2 (y_{i+1}-y_{i-1})$. In principle, the three values $\alpha_x$,
$\alpha_y$, $\alpha_z$ should be identical, but the numerical
accuracy using a second-order differentiation scheme is most
handicapped for those loop segments with the smallest values of
the $B$-component (appearing in the denominator), for instance
in the $\alpha_z$ component $\propto (1/B_z)$ near the loop tops
(where $B_z \approx 0$). It is therefore most advantageous to use
all three parameters $\alpha_x$, $\alpha_y$, and $\alpha_z$ in
a weighted mean,
\begin{equation}
        \alpha = {\alpha_x w_x + \alpha_y w_y + \alpha_z w_z
        \over {w_x + w_y + w_z}} \ ,
\end{equation}
but weight them by the magnitude of the (squared) magnetic field
strength in each component,
\begin{equation}
        w_x = {B_x^2} \ , \quad
        w_y = {B_y^2} \ , \quad
        w_z = {B_z^2} \ ,
\end{equation}
so that those segments have no weight where the $B$-component
approaches zero. This numerical method was found to render the
$\alpha$-parameters most accurately (Aschwanden and Malanushenko 2012),
while the analytical approximation (Eq.~13) breaks down near loop tops
(where $B_z \approx 0$).

The computation of the current densities ${\bf j}(x, y, z)$ follows then
directly from Eq.~(5), 
\begin{equation}
        {\bf j}(x, y, z)/c = \alpha(x, y, z) {\bf B}(x, y, z) \ .
\end{equation}
For instance, a photospheric (vertical) current map, i.e., $j_z(x,y,z_{ph})$, 
as shown in in Fig.~1 (bottom panel), can be calculated with Eq.~(22) for
the photospheric level at $z_{ph}=\sqrt{(1-x^2-y^2)}$. 

\section{       	OBSERVATIONS AND RESULTS 	}

We present now the results from four active regions, which all have been
modeled with different magnetic field models before also 
(DeRosa et al.~2009; Sandman et al.~2009; Aschwanden and Sandman 2010; 
Sandman and Aschwanden 2011) and can now be
compared with our modeling results. A summary of the four active regions,
including the observing time, heliographic position, STEREO spacecraft
separation, number of stereoscopically triangulated loops, 
maximum magnetic field strength, and magnetic flux is given in Table 1,
which is reproduced from Aschwanden and Sandman (2010). 

\subsection{	3D Potential Magnetic Field 	}

The first step of our analysis is the decomposition of the observed
{\sl SOHO}/MDI magnetograms of four active regions into $N_m$ magnetic charges 
(Eq.~2), which yields the $4\times N_m$ parameters $[B_j, x_j, y_j, z_j]$,
$j=1,...,N_m$ for the 3D parameterization ${\bf B}({\bf x})$ of the magnetic
field. In the remainder of the paper we will label the four active regions 
observed at the given times as follows (see also Table 1): 
\begin{verse}
$(A)$ for active region 10953, 2007-Apr-30, 23:00 UT, \\
$(B)$ for active region 10955, 2007-May-9,  20:30 UT, \\
$(C)$ for active region 10953, 2007-May-19, 12:40 UT, \\
$(D)$ for active region 10978, 2007-Dec-11, 16:30 UT. \\
\end{verse}
The pixel size of MDI magnetograms is $2{\arcsec}$ and the 
MDI magnetograms have been decomposed into Gaussian-like peaks 
(each one corresponding to a buried unipolar magnetic charge). The chosen 
field-of-views (FOV) of the four active regions are (in units of solar radii):
\begin{verse}
$FOV(A)=[x_1=-0.65, x_2=+0.05, y_1=-0.50, y_2=+0.20]$, \\
$FOV(B)=[x_1=-0.55, x_2=-0.25, y_1=-0.25, y_2=+0.05]$, \\
$FOV(C)=[x_1=-0.22, x_2=+0.18, y_1=-0.15, y_2=+0.25]$, \\
$FOV(D)=[x_1=-0.28, x_2=+0.12, y_1=-0.35, y_2=+0.05]$. \\
\end{verse}
So, all four active regions are located near disk center $[x=0, y=0]$,
but some extend out to 0.65 solar radii (FOV A).
We show the observed magnetograms $B_z(x,y)$ for the 4 active regions
in Fig.~4 (left column), the corresponding model maps $B_z(x,y)$ 
built from $N_m=200$
magnetic charge components (Fig.~4, middle column), and the difference between
the observed and model maps (Fig.~4, right column), all on the same grey scale,
so that the fidelity of the model can be judged. The residual fields of the
decomposed magnetograms after subtraction of $N_m=200$ components are
\begin{verse}
$B_z/B_{max}=-0.0003\pm0.0097$ for $(A)$, \\
$B_z/B_{max}=-0.0011\pm0.0074$ for $(B)$, \\
$B_z/B_{max}=-0.0004\pm0.0164$ for $(C)$, \\
$B_z/B_{max}=-0.0008\pm0.0189$ for $(D)$, \\
\end{verse}
(with the maximum field strengths $B_{max}$ listed in Table 1). Therefore,
our decomposition algorithm (Appendix A) represents the observed magnetic
field down to a level of $\approx 1\%-2\%$ residuals. 
Note that this method yields
the 3D components of the magnetic field, $[B_x, B_y, B_z]$, with the 
accuracy specified for the case of a potential field, while the accuracy 
for a non-potential field components (i.e., horizontal components
$B_y$ and $B_z$) probably does not exceed a factor of two of this
accuracy (i.e., $\approx 2\%-4\%$), which is still better than the
accuracy of currently available vector magnetograph data (in the order
of $\approx 10\%$ for strong fields and worse for weak fields; 
Marc DeRosa, private communication). 

\subsection{	Forward-Fitting of 3D Force-Free Field 		}

In Fig.~5 we show the comparison of $N_L=200$ stereoscopically triangulated
coronal loops for active region $A$ (2007 Apr 30),
obtained from STEREO spacecraft EUVI/A and EUVI/B data,
with the potential-field model composed from $N_m=100$ buried magnetic
charges (Eq.~2). For each modeled magnetic field line we have chosen the
midpoint of the observed STEREO loop segment as the starting point, from which
we extrapolate the potential field in both directions for segments 
(red curves) that are equally long as the observed loop segments (blue curves). 
A histogram of the median misalignment angles is shown in Fig.~5 (bottom),
measured at 10 positions along each loop segment, which has an average (A) 
of $\mu=39.9^\circ$, a median (M) of $\mu=29.7^\circ$, or a Gaussian peak
(P) fit with a centroid and width of $\mu=29.1\pm21.6^\circ$. So, there
is broad distribution of misalignment angles in the range of $\mu \approx
10^\circ-60^\circ$, which clearly indicates that a potential field model
is not a very good fit for these observables.
  
In Figs.~6-9 we present the main results of our force-free field 
forward-fitting code applied to the stereoscopically triangulated coronal 
loops data, for all four active regions $A$, $B$, $C$, and $D$. 
The 3D coordinates of the observed stereoscopic loops 
(Figs.~6-9; blue curves) and forward-fitted theoretical field lines 
(Figs.~6-9: red curves) are shown in three projections, 
in the $x-y$ plane (in direction of the line-of-sight), as well as in
the orthogonal $x-z$ and $z-y$ planes. 
The theoretical field lines are extrapolated
to equally long segments as the observed EUV loops.
The field lines have been computed with a step size of $\Delta s= 0.002$ 
solar radii (i.e., $\approx 1500$ km), which is identical to 
the spatial resolution of the magnetogram from SOHO/MDI 
($2.0{\arcsec} \approx 1500$ km), and commensurable with the spatial 
resolution of STEREO/EUVI (i.e., 2 pixels with a size of $1.6{\arcsec}$, 
which is $\approx 2000$ km).

Fig.~5 and Fig.~6 show a direct comparison of modeling potential and
non-potential fields for active region $A$, 
which results into a broad Gaussian distribution of 
$\mu=29.1^\circ\pm21.6^\circ$ for the potential field, and to a much
narrower distribution of $\mu=16.7^\circ\pm6.7^\circ$ for the non-potential 
field model. A {\bf online-movie} that visualizes the forward-fitting 
process from the
initial guess of a potential field model to the best-fit non-potential
model is also included in the {\bf online electronic supplementary material}
of this paper. 

\subsection{	Misalignment Statistics 		}

A quantitative measure of the agreement between a theoretical magnetic
field model and the 3D coordinates of observed coronal loops is the
statistics of misalignment angles, which is shown for all four active 
regions in the bottom panels of Figs.~6-9. We find the following
Gaussian distributions of misalignment angles for each of the four active
regions" 
\begin{verse}
$\mu=16.7^\circ \pm 6.7^\circ$ for $A$ (2007 April 30; Fig.~6),\\ 
$\mu=16.0^\circ \pm 6.0^\circ$ for $B$ (2007 May 9; Fig.~7),\\
$\mu=19.2^\circ \pm 6.7^\circ$ for $C$ (2007 May 19; Fig.~8),\\
$\mu=14.3^\circ \pm 5.6^\circ$ for $D$ (2007 Dec 11; Fig.~9),\\ 
\end{verse}
so they have a most frequent value of $\mu \approx 15^\circ-19^\circ$,
which matches very closely the estimated uncertainty of stereoscopic
errors (based on the parallelity of loops in near proximity), which was 
found in a similar range of $\mu_{SE} \approx 7^\circ-10^\circ$ 
(Aschwanden and Sandman 2010). 
{\bf Online-movies} that visualizes the forward-fitting 
process from the initial potential field model to the best-fit non-potential
model are also included in the {\bf online electronic supplementary material}
of this paper.  

A comparison of our results with previous
magnetic modeling of the same four active regions is compiled in Table 2.
The improved misalignment statistics obtained with our code in the range of 
$\mu \approx 14^\circ-19^\circ$ is about two times smaller
than what was obtained with earlier NLFFF models 
($\mu[NLFFF] \approx 24^\circ-44^\circ$; DeRosa et al.~2009),
and also about two times smaller than potential field models, i.e., with
potential source surface models (PFSS:
$\mu \approx 19^\circ-36^\circ$, Sandman et al.~2009),
with potential field models with unipolar charges 
($\mu \approx 16^\circ-26^\circ$, Aschwanden and Sandman~2009),
also measured with our new code described in this paper 
($\mu \approx 18^\circ-42^\circ$). Thus, we find that 
our non-potential (force-free) field model clearly yields a better 
matching model than any previous potential or non-potential field model. 

The mean value of the misalignment angle $\mu$ in an active region 
depends also on the number $N_m$ of magnetic charges that have been
used in the model, which is shown in Fig.~10 (right panels). 
We repeated the forward-fitting for $N_m=1, 2, 4, 10, 20, 50, 100, 200$
magnetic charges and find that the mean misalignment angle $\mu(N_m)$ 
generally improves (or decreases) with the number of magnetic
components, up to $N_m\approx 100$, while for larger numbers (i.e.,
$N_m=200$) a diminuishing effect sets in (probably due to less efficient
convergence in forward-fitting with a too large number of variables).
We show the dependence of the mean misalignment angle $\mu(N_m)$ for
both the potential field model (P) and the force-free field non-potential
model (NP) in Fig.~10 (right panels), which reveal interesting
characteristics how potential-like an active region is. Clearly,
active region B (2007 May 9) is the most potential one, while
active region C (2007 May 19) is the most non-potential one, where
a potential field model does not fit at all, regardless how many
magnetic components are used. A similar result was also obtained in
the study of Aschwanden and Sandman (2010), where the potential-like
active region B was associated with the lowest GOES-class level (A7),
while the most non-potential-like region C was exhibiting a GOES-class
C0 flare. 

\subsection{Force-Free $\alpha$ and Electric Current $j_z$ Maps} 

Maps of the magnetic field components $B_x(x,y)$, $B_y(x,y)$, and
$B_z(x,y)$ obtained from the decomposition of line-of-sight magnetograms
into 100 buried magnetic charges are shown for one (i.e., region A
of 2007 Apr 30) of the four analyzed active regions in Fig.~11. 
Note that only the line-of-sight component map $B_z(x,y)$ is directly
observed (with SOHO/MDI), as shown in the bottom right panel in Fig.~11,
while the other two component maps $B_x(x,y)$ and $B_y(x,y)$ are 
inferred from the spherical symmetry of the magnetic field of point charges
assumed for a potential field (Eq.~1). These inferred field components
define our potential field solution ${\bf B}^{P}(x,y)$.

Our non-potential field model adds additional azimuthal field components
$B_{\varphi}(x,y)$ that are only constrained by fitting the 
stereoscopically triangulated loops. We show the difference
between the non-potential and potential field components for the
photospheric level in the middle column of Fig.~11. We notice that the 
difference
between the non-potential and potential solution is essentially contained
in the transverse field components $B_x(x,y)$ and $B_y(x,y)$,
because the line-of-sight component $B_z(x,y)$ is an observed 
quantity and has to be matched by any model. The difference maps shown
in Fig.~11 confirm that our model retrieves the non-potential field
mostly from the transverse field components that are not measured
with a line-of-sight magnetogram. 

From our model we can also derive a photospheric force-free $\alpha(x,y)$ 
map (top right panel in Fig.~11) and a photospheric electric current 
density map $j_z(x,y)$ (middle right panel in Fig.~11), based on Eq.~22.
The magnetograms and inferred current maps are
shown for active region (A) in Fig.~11, and for the other active regions
(B), (C), and (D) in Fig.~12. These inferred maps reveal that the 
locations of significant electric currents are located in strong field
regions, and exhibit as much small-scale fine structure as the magnetic
field line-of-sight component $B_z(x,y)$.

Statistics of obtained parameters of our non-potential field model
is shown in form of histogrammed distributions in Fig.~13, lumped
together from all 454 stereoscopically triangulated loops of our 
four analyzed active regions. The histograms are shown in lin-lin
(Fig.~13 left) as well as in log-log representation (Fig.~13 right) and
can be characterized by powerlaw distributions.
The median values are: $L=163$ Mm for the length of the extrapolated
field lines that were extrapolated through the midpoint of the observed
loops, 
$N_{twist}=0.06$ for the number of twisted turns along the extrapolated
full loop lengths; $\alpha = 4 \times 10^{-11}$ cm$^{-1}$ for the
force-free $\alpha$-parameter, and $|j_z| = 1500$ Mx cm$^{-2}$ s$^{-1}$
for the electric current density. The statistics of misalignment
angles is given in Fig.~10 and Table 2.

\subsection{	Divergence-Freeness and Force-Freeness	}

In order to test the accuracy of our analytical force-free model,
which represents an approximation to a truly force-free field with
an accuracy to second order (in $\alpha$), it is useful to calculate 
some figures of merit. The divergence-freeness $\nabla \cdot {\bf B} = 0$ 
can be compared with the field gradient $B/\Delta x$ over a pixel length 
$\Delta x$,
\begin{equation}
        L_d = {1 \over V} \int_V
        {|(\nabla \cdot {\bf B}) |^2
        \over |B / \Delta x|^2} dV \ .
\end{equation}
Similarly, the force-freeness can be quantified by the ratio of the
Lorentz force, $({\bf j} \times {\bf B}) = (\nabla \times {\bf B}) \times
{\bf B}$ to the normalization constant $B^2 / \Delta x$,
\begin{equation}
        L_f = {1 \over V} \int_V
        {|(\nabla \times {\bf B}) \times {\bf B}|^2
        \over |B^2 / \Delta x|^2}  dV \ ,
\end{equation}
where $B = |{\bf B}|$. These quantities, integrated over a computational
box that covers the field-of-views of an active region and extends over
a height ragne of $\Delta h = 0.15$ solar radii, were found to be for the
four active regions A, B, C, and D:
\begin{verse}
	(A) $L_d=0.5 \times 10^{-4}$ and $L_f=5 \times 10^{-4}$ \\
	(B) $L_d=1.1 \times 10^{-4}$ and $L_f=7 \times 10^{-4}$ \\
	(C) $L_d=1.4 \times 10^{-1}$ and $L_f=1.8 \times 10^{-1}$ \\
	(D) $L_d=1.0 \times 10^{-4}$ and $L_f=1.0 \times 10^{-4}$ \\
\end{verse}
In comparison, figure of $L_d \approx (8 \pm 5) \times 10^{-4}$ and
$L_f \approx (24 \pm 23) \times 10^{-4}$ were found for simulated cases
(Aschwanden and Malanushenko 2012), and 
$L_d \approx (0.2 \pm 8) \times 10^{-4}$ and
$L_f \approx (6 \pm 2) \times 10^{-4}$ for forward-fitting to the
Low and Lou (1990) model, which is an analytical exact force-free solution
(Aschwanden and Malanushenko 2012). Our values do not exceed figures of
merits quoted from other NLFFF codes, i.e., $L_d \approx 0.07$ and 
$L_f \approx 0.1$ (Schrijver et al.~2006; Table III therein). For sake
of convenience we evaluated these figures of merit in a rectangular
box tangential to the solar disk at disk center, extending over a height 
range of $h=1.00-1.15$, which covers the strong field regions
only for active regions near disk center. Of course, the widely used
definition of divergence-freeness (Eq.~23) and force-freeness (Eq.~24)
is proportional to the square of the normalization length scale $L$, i.e.,
$L_d \propto L^2$ and $L_f \propto L^2$, and thus would yield larger
values for typical loop lengths or box sizes $L \gg \Delta x$.

\section{       	DISCUSSION 			}

\subsection{	The Potential Field Model with Unipolar Magnetic Charges }

As a first step we derived a potential field model of an active region.
The knowledge of the potential field is a useful starting point
for reconstructing a non-potential field model. In the asymptotic
limit of small currents, i.e., for small values of the force-free 
parameter ($\alpha \mapsto 0$), the force-free field solution automatically 
converges to the potential field solution. Furthermore, since our analytical
force-free field approximation is accurate to second order (in $\alpha$),
the highest accuracy is warranted for a small non-potentiality, say up
to one full turn of twist along a loop length. The fact that we measured
small twists with a statistical median of $N_{twist} \approx 0.06$ for
the four analyzed active regions, justifies the neglect of second- and
higher-order terms in our analytical approximation.

Generally, vector magnetograph data are required to uniquely define 
the complete 3D magnetic field ${\bf B}({\bf x})$ at the photospheric 
boundary. In constrast, potential-field codes can extrapolate a unique
3D field solution from a single B-component only, i.e., from the 
line-of-sight component $B_z({\bf x})$. Consequently,
in our method of superimposed fields from buried magnetic charges,
the potential field solution is unique in every point of space, as well as
at the boundaries (neglecting exterior magnetic charges). The potential-field
solution fixes also the line-of-sight component of the non-potential
solution, $B_z^{NP} = B_z^{P}$, because this component is a direct
observable that must be matched with every model, and thus is identical
for both the potential and the non-potential model, while the transverse 
components represent the only free parameters for a non-potential
solution, i.e., $B_x^{NP} \ne B_x^{P}$ and $B_y^{NP} \ne B_y^{P}$. 

In this study we derived an algorithm that accurately calculates 
a potential magnetic field by properly including the curvature of the solar
surface and for locations away from the solar disk center. We are not aware 
that such a potential-field code, defined in terms of buried unipolar magnetic
charges and deconvolved from an observed magnetogram (Aschwanden and 
Sandman 2010), has been developed elsewhere, although similar parameterizations
have been used in {\sl magnetic charge topology (MCT)} models
(e.g., Longcope 2005). It would be 
interesting to compare its performance with other potential field codes, 
such as with the Green's function method (Sakurai 1982), the eigenfunction
expansion method (Altschuler and Newkirk 1968), or the potential field
source surface (PFSS) code (e.g., Luhmann et al.~1998). 
Our magnetic charge decomposition method is related to the Green's function 
method, but differs in the discretization of discrete magnetic elements of
various strengths and variable location, while the Green's method uses
a regular surface grid. Magnetic elements with sub-pixel size represent
no particular problem (except that no more than one element can be resolved
per pixel), because our method fits their radial field above the photosphere,
but treats them as point charges below the photosphere. However, 
the resolution of our method is somehow limited by the depth of the point 
charges (which introduces finite numerical errors in the calcuation of
$\nabla \cdot {\bf B}$ and $\nabla \times {\bf B}$), as well as by
the number of point charges (which represent a sensitivity threshold 
for weak magnetic field sources). Thus, smaller grid pixel sizes and
larger number of magnetic field charges can enhance the accuracy of
the solutions, but are more demanding regarding computation times.
Similarly, the resolution of the Green's function method is limited 
by the grid size, which is either given by the measurements, or limited 
by computation ressources. 

\subsection{	Non-Potential Fields with Twisted Loops		}

Models of twisted flux tubes have been applied abundantly in solar physics,
e.g., to braiding of coronal loops (Berger 1991), to prominences (Priest et al. 
1989, 1996), to sigmoid-shaped filaments (Rust and Kumar 1996; 
Pevtsov et al.~1997), to emerging current-carrying flux tubes (Leka et 
al.~1996; Longcope and Welsch 2000), or to turbulent coronal heating 
(Inverarity and Priest 1995). Magnetic structures are believed to be
twisted and current-carrying before they emerge at the solar surface
(Leka et al.~1996), many active region loops are observed to have a
visible twist (see Section 6.2.4 in Aschwanden 2004), 
and filaments or prominences become unstable and erupt
once their twist exceeds a critical angle of a few full turns due to
the kink instability or torus instability (Fan and Gibson 2003, 2004;
T\"or\"ok and Kliem 2003; Kliem et al.~2004). The twist of a magnetic
structure is therefore an important indicator for its stability or
transition to an instability, followed by the dynamic evolution that
leads to eruptive flares and coronal mass ejections (CME).
Our method allows us to determine the number of twisting turns 
of most stereoscopically triangulated coronal loops directly, and thus 
provides a reliable diagnostics on its stability and the amount of
electric current the loop carries. The statistics in Fig.~13 shows
that the number of twisting turns has a range up to $N_{twist} 
\lapprox 0.25$ (Fig.~13), which is far below the critical limit of 
$N_{twist}\approx 1.2-2.4$ full turns required for the onset of the 
kink instability in a force-free magnetic field (Mikic et al.~1990;
T\"or\"ok and Kliem 2003).

The amount of twisting clearly varies among different active regions.
The improvement in the misalignment angle between a potential and a
non-potential model characterizes the potentiality and free energy
of an active region. The most potential active region is B (2007 May 9),
where we achieve only a slight improvement of $\Delta \mu
=\mu^P-\mu^{NP}=19^\circ-16^\circ=3^\circ$ (Table 2, for the Gaussian
peak of the distribution). Also active region D (2007-Dec-11) is moderately
potential, where we find $\Delta \mu=\mu^P-\mu^{NP}=21^\circ-15^\circ=6^\circ$.
Significant non-potentiality is found for active region A (2007-Apr-30),
where the improvement amounts to
$\Delta \mu =\mu^P-\mu^{NP}=29^\circ-17^\circ=12^\circ$. The strongest
non-potentiality is found for active region C (2007-May-19), where we
achieve an improvement of 
$\Delta \mu=\mu^P-\mu^{NP}=46^\circ-20^\circ=26^\circ$.
Therefore, the misalignments reduce by an amount of $\Delta \mu=3^\circ
..., 26^\circ$
for these four active regions. During the time of observations,
active region C indeed featured a GOES-class C0 flare, which explains
its non-potentiality. The remaining amount of misalignment, in the order
of $\mu=15^\circ-20^\circ$ is attributed partly to
stereoscopic measurement errors, which were estimated to $\mu_{SE} \approx 
8^\circ-12^\circ$, and partly to our particular parameterization of
a force-free field model, which is optimally designed for vertically twisted
structures. 

\subsection{	Force-Free $\alpha$ Parameter and Electric Currents  }	

For the force-free $\alpha$ parameter we find a distribution extended
over the range of $\alpha \lapprox 20 \times 10^{-11}$ cm$^{-1}$
(Fig.~13), with a median of $\alpha \approx 4 \times 10^{-11}$ cm$^{-1}$.
Using a sheared arcade model, where the force-free $\alpha$ parameter
is defined as $\alpha=L \tan \mu$ (with $L$ the length of the loops),
a range of $\alpha = (0.6-13) \times 10^{-11}$ cm$^{-1}$ was found
for the same dataset, based on the residual misalignment of
$\mu \approx 7^{\circ}-13^{\circ}$ attributed to non-potentiality 
(Table 4 in Sandman and Aschwanden 2011). Modeling of stereoscopically
triangulated loops with a linear force-free model yielded also a similar
range of $\alpha \approx (2-8) \times 10^{-11}$ cm$^{-1}$
(Feng et al.~2007b). Thus, the range of our determined $\alpha$
parameters agrees well with these three studies.

The current densities have been determined in a range of
$j \approx 10^{-2}-10^4$ Mx cm$^{-2}$ s$^{-1}$, with a median of
$j \approx 1500$ Mx cm$^{-2}$ s$^{-1}$ (Fig.~13), for spatial locations
with $B>100$ G at the photospheric level. For comparison,
Leka et al.~(1996) determined similar currents of
$j_z \approx (21-75)$ Mx cm$^{-2}$ s$^{-1}$ (i.e., $7-25$
mA m$^{-2}$, see Table 3 in Leka et al.~1996), using vector magnetograph
data at the Mees Solar Observatory measured in emerging bipoles in
active regions. In principle, the knowledge of the magnitude of
the electric current density $j$ allows one to estimate the amount of
Joule dissipation,
\begin{equation}
	E_{H} = {j^2 \over \sigma} \ ,
\end{equation}
where $\sigma \approx 6 \times 10^{16}$ s$^{-1}$ is the classical conductivity
for a $T\approx 2$ MK hot corona. Our measured maximum current density
of $|j_z| < 10^4$ Mx cm$^{-2}$ s$^{-1}$ yields then a value of
$E_H < 2 \times 10^{-9}$ erg cm$^{-3}$ s$^{-1}$ for the volumetric
heating rate. The corresponding
Poynting flux $F_H$ for a field line with a density scale height of
$\lambda \approx 10^{10}$ cm (corresponding to the thermal scale height
at a temperature of $T=2$ MK), is 
\begin{equation}
	F_H = E_H \ \lambda(T) \approx 10^{10} E_H 
		\left({T_e \over 2\ {\rm MK}}\right) \ , 
\end{equation}
which yields $F_H < 20$ erg cm$^{-2}$ s$^{-1}$,
which is far below the heating requirement for active regions
($F_H \approx 10^5-10^7$ erg cm$^{-2}$ s$^{-1}$; Withbroe and Noyes 1977).
Thus, Joule dissipation is far insufficient to heat coronal loops in
active regions, according to our current measurements. Obviously,
more energetic processes, such as magnetic reconnection with anomalous
resistivity in excess of classical conductivity is needed.
Nevertheless, our technique of measuring current densities from the
twist of observed loops may provide a useful diagnostic where currents
are generated in active regions, e.g., near neutral lines with large
gradients in the magnetic field, or in areas with large photospheric
shear that produces magnetic stressing.

\subsection{	Benefits of Non-Potential Magnetic Field Modeling 	}	

Modeling of the magnetic field with a non-potential model 
matches the observed loop geometries about a 
factor of two better than potential field models, as demonstrated 
in this study (Table 2).
While our attempt of forward-fitting of twisted fields to 
stereoscopically triangulated loops represents
only a first step in this difficult problem of data-constrained 
non-potential field modeling, we expect that future non-potential models
with similar parameterizations will be developed, which yield the
3D magnetic field with high accuracy, satisfying Maxwell's
equations of divergence-freeness and force-freeness.
There are a number of benefits that will result from the knowledge 
of a realistic 3D magnetic field, of which we just mention a few:

\begin{enumerate}
\item{The spatial distribution of coronal DC current heating can be 
	inferred from localizations of non-potential field lines 
	with a high degree of twist and current density (e.g., 
	Bobra et al.~2008; DeRosa et al.~2009; Su et al.~2011).}
\item{The free (non-potential) energy $E_{free}=E_{N}-E_P$ can be
	determined from the difference of the non-potential ($E_N$) and
	potential ($E_P$) energy in a loop.}
\item{Correlations between the free energy $E_{free}$ and the
	soft X-ray brightness $I_{SXR}$ can quantify the 
	volumetric heating rate $E_H$ and pointing flux $F_H$
	in loops and active regions, and be related to the
	flare productivity (e.g., Jing et al.~2010; 
	Aschwanden and Sandman 2010).}
\item{The helicity and its evolution can be traced in twisted
	coronal loops, which can be used to diagnose whether
	helicity injection from below the photosphere takes place
	(e.g., Malanushenko 2011b).}
\item{Hydrodynamic modeling of coronal loops, which is mostly done
	in a one-dimensional coordinate $(s)$ along the loop, 
	requires the knowledge of the 3D geometry $[x(s), y(s), z(s)]$,
	which otherwise can only be obtained from stereoscopy.
	The inclination of the loop plane determines the ratio
	of the observed scale height to the effective thermal 
	scale height, and thus the inferred temperature profile
	$T_e(s)$. Also the inference of the electron density
	$n_e(s)$ along the loop depends on the column depth of the
	line-of-sight integration (e.g., Aschwanden et al.~1999).
	Most loop scaling laws depend explicitly on the loop length $L$
	(e.g., the Rosner-Tucker-Vaiana law, Rosner et al.~1978),
	which can only be reliably determined from magnetic models.}
\item{Electron time-of-flight measurements require the trajectory
	of electrons that stream along field lines from coronal
	acceleration sources to the photospheric footpoints. 
	The location of the coronal acceleration can be 
	determined from the observed energy-dependent time delays
	in hard X-rays and the knowledge of the pitch angle of the 
	electron and the magnetic twist of the field line, which can 
	only be obtained from magnetic models (e.g., Aschwanden et al.~1996).}
\item{Modeling of entire active regions require models of the 3D magnetic 
	field, the temperature $T_e(s)$, and density $n_e(s)$, using 
	hydrostatic or hydrodynamic loop models, which can reveal 
	scaling laws between the volumetric heating rate $E_H$, 
	hydrodynamic loop parameters ($n_e, T_e, p, L$), and magnetic
	parameters ($B$), (e.g., Schrijver et al.~2004;
	Warren \& Winebarger 2006, 2007; Warren et al.~2010;
	Lundquist et al.~2008a,b).}
\item{Nonlinear force-free modeling of active regions and global coronal
	fields can establish better lower boundary conditions for 
	modeling of the heliospheric field (e.g., Petrie et al. 2011).}
\end{enumerate}

\section{       	CONCLUSIONS 			}

There are two classes of coronal magnetic field models, potential fields 
and non-potential fields, which both so far do not fit the observed 3D
geometry of coronal loops well. Nonlinear force-free fields (NLFFF) are
found to match the data generally better than linear force-free fields (LFFF).
The 3D geometry of coronal loops is most
reliably determined by stereoscopic triangulation, as it is now available
from the twin STEREO/A and B spacecraft, and we made use of a sample of
some 500 loops observed in four active regions during the first year
of the STEREO mission (with spacecraft separation angles in the range
of $\alpha_{sep} \approx 6^\circ-43^\circ$). We forward-fitted a 
force-free approximation to the entire ensemble of stereoscopically
triangulated loops in four active regions and obtained the following
results and conclusions:

\begin{enumerate}
\item{A line-of-sight magnetogram that measures the longitudinal
	magnetic field component $B_z(x,y)$ can be decomposed into
	$N_m\approx 100$ unipolar magnetic charges, from which maps
	of all three magnetic field components $B_x(x,y)$, $B_y(x,y)$,
	and $B_z(x,y)$ can be reconstructed at the solar surface with
	an accuracy of $\approx 1\%-2\%$. This boundary condition
	allows us to compute a 3D potential field model ${\bf B}^P({\bf x})$
	in a 3D cube encompassing an active region, by superimposing the
	potential fields of each buried magnetic charge. Our algorithm
	takes the curvature of the solar surface into account and is 
	accurate up to about a half solar radius away from disk center.}
\item{Forward-fitting of the twisted flux tube model to 70-200 loops 
	per active region improves the median 3D misalignment angle 
	between the theoretical field lines and the observed stereoscopically
	triangulated loops from $\mu=19^\circ-46^\circ$ for a potential field
	model to $\mu=14^\circ-19^\circ$ for the non-potential field model,
	which corresponds to a reduction of $\Delta \mu = \mu^P-\mu^{NP}
	=3^\circ,...,26^\circ$. The residual misalignment is commensurable with 
	the estimated stereoscopic measurement error of $\mu_{SE} 
	\approx 8^\circ-12^\circ$.}
\item{The application of our stereoscopy-constrained model allows us to
	obtain maps of the non-potential magnetic field components,
	$B_x(x,y), B_y(x,y), B_z(x,y)$, the force-free $\alpha$-parameter 
	$\alpha(x,y)$, and the current density $j_z(x,y)$ at the photospheric
	level or in an arbitrary 3D computation box.
	The divergence-freeness and force-freeness, numerically evaluated
	over a 3D computation box, was found to be reasonable well fulfilled.} 
\item{The statistics of parameters obtained from forward-fitting of our
	force-free model yields the following values:
	field line lengths $L \approx 50-300$ Mm (median 163 Mm),
	number of twist turns $N_{twist} \lapprox 0.25$ 
	(median $N_{twist}=0.06$), nonlinear force-free $\alpha$-parameter 
	$\alpha \lapprox 15 \times 10^{-11}$ cm$^{-1}$
	(median $\alpha \approx 4 \times 10^{-11}$ cm$^{-1}$),
	and current density $|j_z| \approx 10^{-2}-10^4$ Mx cm$^{-2}$ s$^{-1}$
	(median $|j_z| \approx 1500$ Mx cm$^{-2}$ s$^{-1}$).
	All twisted loops are found to be far below the critical value
	for kink instability ($N_{twist} \approx 1.25$ turns) and
	Joule dissipation of their currents (with a median Poynting flux
	of $F_H < 20$ erg cm$^{-2}$ s$^{-1}$) is found be be far below
	the coronal heating requirement ($F_H \approx 10^5-10^7$ 
	erg cm$^{-2}$ s$^{-1}$).}
\end{enumerate}
	
Where do we go from here? The two algorithms developed here provide
an efficient tool to quickly compute a potential field and a quasi-force-free
solution of an active region, yielding also accurate measurements of geometric 
(helically twisted) loop parameters (full length of field line, number of 
twist turns), based on the constraints of stereoscopically triangulated loops.
Our approach of forward-fitting a magnetic field model to coronal structures
has also the crucial advantage to bypass the non-force-free zones in the lower
chromosphere, which plague standard NLFFF extrapolation algorithms.
A next desirable project would be to
generalize the non-potential field forward-fitting algorithm to
2D projections $[x(s), y(s)]$ of 3D loop coordinates $[x(s), y(s), z(s)]$ 
(as they are obtained from stereoscopic triangulation), 
so that the model works for a
single line-of-sight magnetogram combined with a suitable 
single-spacecraft EUV image, without requiring stereoscopy.
The stereoscopic 3D loop measurements (used here) as well as
3D vector magnetograph data (once available from HMI/SDO), however,
represent important test data to validate any of these non-potential
field models. 

\section*{APPENDIX A: Deconvolution of Magnetic Charges }

A 3D parameterization of a line-of-sight magnetogram $B_z(x,y)$ can be
obtained by a superposition of buried magnetic point charges, which produce
a surface magnetic field ${\bf B}=(B_x, B_y, B_z)$ that is constrained by
the observed magnetogram, as defined in Section 2.1. Here we describe the
geometric inversion of the 3D coordinates $(x_m, y_m, z_m)$ and surface
field strength $B_m$ of a single magnetic point charge $M$ that produces 
a local peak $B_z$ with width $w$ at an observed position 
$(x_p, y_p)$ (Fig. 14, left). The geometric relationships
can be derived most simply in a plane that intersects the point souce $P$
and the line-of-sight axis. In the plane-of-sky we define a coordinate axis
$\rho$ that is orthogonal to the line-of-sight axis $z$ and is rotated
by an angle $\gamma$ with respect to the $x$-axis,
so that we have the transformation,
$$
	\begin{array}{ll}
        \rho_p =&\sqrt{x_p^2+y_p^2} \\
	x_p =& \rho_p \ \cos(\gamma) \\
	y_p =& \rho_p \ \sin(\gamma) \\
        z_p  =& \sqrt{ 1 - \rho_p^2} \\
	\end{array} \ .
	\eqno(A1) 
$$	
Thus we have the four observables $(B_z, \rho_p, z_p, w)$ and want to
derive the model parameters $(B_m, x_m, y_m, z_m)$. In Fig.~14 (right
hand side) we show the geometric definitions of the depth $d_m$ of
the magnetic charge (the radial distance between $M$ and the surface), 
the distance $d$ between the magnetic charge $M$ and the surface field
at an observed position $P$, which is inclined by an 
angle $\beta$ to the vertical direction above the magnetic charge $M$,
so we have the relation,
$$
	\cos{ \beta} = {d_m \over d} \ .		
	\eqno(A2)
$$
The line-of-sight component $B_z$ at point $P$ has an angle $(\alpha-\beta)$
to the radial direction $B_r$ with field strength $B_r=B_m (d_m/d)^2$
(Eq.~1), and thus obeys the following dependence on the aspect angle $\alpha$
and inclination angle $\beta$ (with A2),
$$
	B_z      = B_m \left( {d_m \over d} \right)^2 \cos{(\alpha - \beta)}
	           = B_m \cos^2{(\beta)} \cos{(\alpha - \beta)} \ .
	\eqno(A3)
$$
The radial coordinate $\rho_p$ is related to the radial coordinate
$\rho_m$ of the magnetic charge by,
$$
	\rho_p = \rho_m + d \sin{(\alpha - \beta)}
		     = \rho_m + d_m {\sin{(\alpha - \beta)} \over \cos{\beta}}
	\ .
	\eqno(A4)
$$
Further we have the geometric relationships for the aspect angle $\alpha$,
the 3D distance $r_m$ from Sun center, and the depth $d_m$, 
$$
	\begin{array}{ll}
	\alpha  =& \arctan{\left( {\rho_m / z_m}\right)} \\
	r_m	=& \sqrt{\rho_m^2 + z_m^2} \\
	d_m     =& (1 - r_m) \\
	\end{array} \ .
	\eqno(A5)
$$
The observed line-of-sight component $B_z$ has a dependence on the inclination 
angle $\beta$ (Eq.~A3), and thus we need to compute the optimum angle
$\beta_p$ where the component $B_z$ has a maximum, because
we can only measure the locations of local peaks in magnetograms
$B_z(x,y)$. We obtain this optimum angle $\beta_p$ by calculating the
derivative $\partial B_z/d\beta$ from Eq.~(A3) and setting the
derivative to zero at the local maximum, i.e.,  
$\partial B_z/d\beta = 0$ at $\beta=\beta_p$, which yields a quadratic
equation for $\tan{(\beta_p)}$ that has the analytical solution,
$$
	\tan{(\beta_p)} = {\sqrt{9 + 8 \tan^2 \alpha} - 3 \over 4 \tan{\alpha}}
	\approx {\alpha \over 3} \ . 
	\eqno(A6)
$$
For the special case of a source at disk center ($\alpha = 0$) this
optimum angle is $\beta_p=0^\circ$, but increases monotonically with
the radial distance from disk center and reaches a maximum value
of $\beta_p = \arctan{(1/\sqrt{2})} \approx 35.26^\circ$ at the limb
($\alpha = \pi/2$). 

Further we need to quantify the half width $w$ of the radial magnetic field
profile $B_z(\rho)$ across a local peak, which is also one of the observables.
A simple way is to approximate the magnetic field profile $B_z{(\rho)}$ with
a Gaussian function, which drops to the half value $B_z/2$ at an angle
$\beta=\beta_2$,
$$
	B_z(\beta=\beta_2) = {1 \over 2} B_z(\beta=\beta_p) \ .
	\eqno(A7)
$$
Inserting the function $B_z(\beta)$ (Eq.~A3) yields then the following
relationship for the half width $w=(\rho_2-\rho_p)$, 
$$
	w = (\rho_2 - \rho_p) = d_m \left[ {\sin{(\alpha-\beta_2)}
	\over \cos \beta_2} - {\sin{(\alpha-\beta_p)} \over \cos \beta_p}
	\right] \ .
	\eqno(A8)
$$
For the special case at disk center $\alpha=0$, the solution is
$w_0=w(\alpha_0)=d_m \tan{(\beta_2)}$, and $\beta_2$ is related to
$\beta_p$ by
$$
	\cos^3{(\beta_2)} = {\cos^3{(\beta_p)} \over 2} \ .
	\eqno(A9)
$$
A general analytical solution is not possible, but numerical inversions of 
Eq.~(A3) give the following very close approximation,
$$
	w \approx d_m \ \tan{\beta_2} \ \cos{\alpha} \ (1 - 0.1 \alpha) \ .
	\eqno(A10)
$$

We have now all geometric relationships to invert the theoretical
parameters $(B_m, x_m, y_m, z_m)$ from the observables 
$(B_z, \rho_\beta, z_\beta, w)$. An explicit derivation of the theoretical
parameters is not feasible, but an efficient way is to start with
an approximate value for the aspect angle, since $tan{(\alpha)}
=(\rho_m/z_m) \approx (\rho_p/z_p)$, followed by a few
iterations to obtain the accurate value. The inversion can be done
in the following order (using Eqs.~A1-A10),
$$
	\begin{array}{ll}
	\alpha  &\approx \arctan({\rho_p / z_p}) \\
 	\beta_p &=\arctan{\left[ \left( \sqrt{9 + 8 \tan^2 \alpha}-3 \right) 
		  / 4\ \tan{\alpha} \right]} \\
	B_m     &={ B_z / [\cos^2{\beta_p} \ \cos{(\alpha-\beta_p)}]} \\
		\beta_2 &=\arccos{\left[ 
		\left( (\cos{\beta_p})^3 / 2 \right)^{1/3} \right]} \\
	d_m     &={w / \left[ \tan{\beta_2}\ \cos{\alpha} \ (1-0.1\alpha) 
		\right]} \\
	r_m	&=(1-d_m)	\\
	\rho_m  &=\rho_p - d_m {\sin{(\alpha-\beta_p)} /
		\cos{\beta_p} } \\
	z_m     &=\sqrt{r_m^2-\rho_m^2} \\
	x_m	&=\rho_m \ \cos{\gamma} \\
	y_m	&=\rho_m \ \sin{\gamma} \\
	\end{array}
	\eqno(A11)
$$
In our algorithm we iteratively determine the local peaks of the line-of-sight
components $B_z(x,y)$ and their widths $w$, which are found to be close to 
Gaussian 2D distribution functions, and invert the model parameters
$(B_m, x_m, y_m, z_m)$ with the inversion procedure given in Eq.~(A11).

\acknowledgements
We thank the anonymous referee and Allen Gary for constructive and helpful comments.
This work is supported by the NASA STEREO under NRL contract N00173-02-C-2035.
The STEREO/ SECCHI data used here are produced by an international consortium of
the Naval Research Laboratory (USA), Lockheed Martin Solar and Astrophysics Lab
(USA), NASA Goddard Space Flight Center (USA), Rutherford Appleton Laboratory
(UK), University of Birmingham (UK), Max-Planck-Institut f\"ur
Sonnensystemforschung (Germany), Centre Spatiale de Li\`ege (Belgium), Institut
d'Optique Th\'eorique et Applique (France), Institute d'Astrophysique Spatiale
(France).
The USA institutions were funded by NASA; the UK institutions by 
the Science \& Technology Facility Council (which used to be the Particle
Physics and Astronomy Research Council, PPARC); the German institutions by
Deutsches Zentrum f\"ur Luft- und Raumfahrt e.V. (DLR); the Belgian institutions
by Belgian Science Policy Office; the French institutions by Centre National
d'Etudes Spatiales (CNES), and the Centre National de la Recherche Scientifique
(CNRS). The NRL effort was also supported by the USAF Space Test Program and
the Office of Naval Research.


\clearpage


\begin{deluxetable}{llllrrcc}
\tabletypesize{\footnotesize}
\tablecaption{Data selection of four Active Regions observed with STEREO/EUVI
and SOHO/MDI.}
\tablewidth{0pt}
\tablehead{
\colhead{\#}&
\colhead{Active}&
\colhead{Observing}&
\colhead{Observing}&
\colhead{Spacecraft}&
\colhead{Number}&
\colhead{Magnetic}&
\colhead{Magnetic}\\
\colhead{}&
\colhead{Region}&
\colhead{date}&
\colhead{times}&
\colhead{separation}&
\colhead{of EUVI}&
\colhead{field strength}&
\colhead{flux}\\
\colhead{}&
\colhead{Active}&
\colhead{}&
\colhead{(UT)}&
\colhead{angle (deg)}&
\colhead{loops}&
\colhead{B(G)}&
\colhead{($10^{22}$ Mx)}}
\startdata
A& 10953 (S05E20)  &2007-Apr-30    &23:00-23:20    &6.0$^\circ$  &200    &[-3134,+1425]& 8.7\\
B& 10955 (S09E24)  &2007-May-9     &20:30-20:50    &7.1$^\circ$  &70     &[-2396,+1926]& 1.6\\
C& 10953 (N03W03)  &2007-May-19    &12:40-13:00    &8.6$^\circ$  &100    &[-2056,+2307]& 4.0\\
D &10978 (S09E06)  &2007-Dec-11    &16:30-16:50    &42.7$^\circ$ &87     &[-2270,+2037]& 4.8\\
\enddata
\end{deluxetable}

\begin{deluxetable}{lllll}
\tablecaption{Misalignment statistics of four analyzed active regions.}
\tablewidth{0pt}
\tablehead{
\colhead{Parameter}&
\colhead{2007-Apr-30}&
\colhead{2007-May-9}&
\colhead{2007-May-19}&
\colhead{2007-Dec-11}}
\startdata
Misalignment NLFFF$^1$& $24^\circ-44^\circ$  &         &       &       \\
Misalignment PFSS$^2$ & $25^\circ \pm  8^\circ$ & $19^\circ \pm 6^\circ$ & 
                        $36^\circ \pm 13^\circ$ & $32^\circ \pm10^\circ$ \\
Potential Field$^3$   & $29^\circ \pm 22^\circ$ & $19^\circ \pm 7^\circ$ 
		      & $46^\circ \pm 23^\circ$ & $21^\circ \pm 9^\circ$ \\
Force-Free Field$^4$  & $17^\circ \pm  7^\circ$ & $16^\circ \pm 6^\circ$ & 
                        $19^\circ \pm  7^\circ$ & $14^\circ \pm 6^\circ$ \\
Stereoscopy error SE$^5$&$9^\circ$ & $8^\circ$ & $12^\circ$ & $9^\circ$ \\
\enddata
\par\noindent $^1)$ NLFFF = nonlinear force-free field code (DeRosa et al.~2009),
\par\noindent $^2)$ PFSS = Potential field source surface code (Sandman et al.~2009),
\par\noindent $^3)$ Potential field model (this work),
\par\noindent $^4)$ Force-free field model (this work).
\par\noindent $^5)$ SE: Measured from inconsistency between adjacent loops.
\end{deluxetable}

\clearpage


\begin{figure}
\plotone{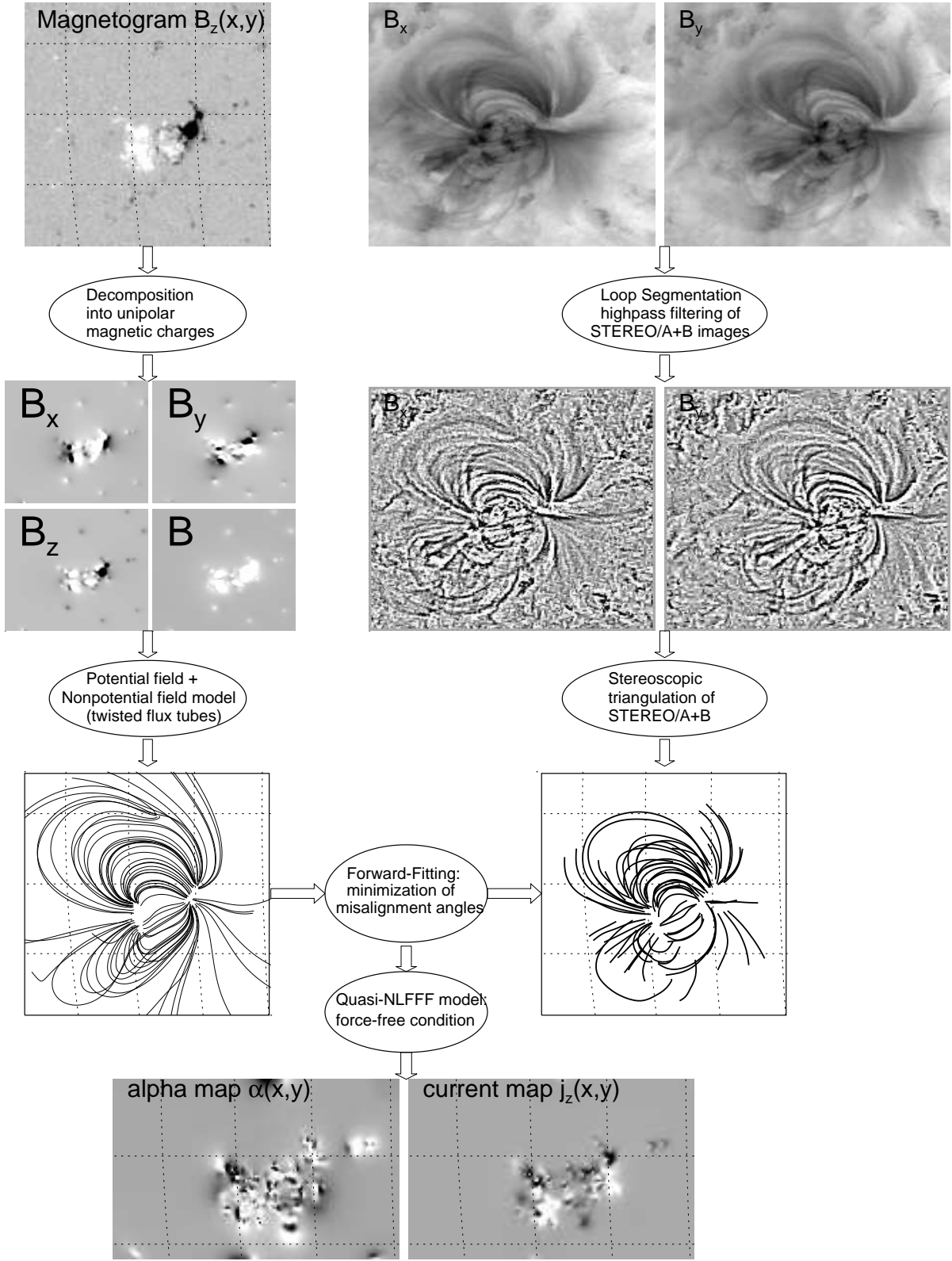}
\caption{Flow chart of magnetic field modeling algorithm: Starting from 
a line-of-sight magnetogram (top left), the 3D magnetic field components 
and field lines of a potential field are calculated. On the other side,
3D loop coordinates are calculated from a pair of STEREO A and B images 
using stereoscopic triangulation (top right). A quasi-force-free magnetic
field model with variable parameters that represent twisted loops is then 
forward-fitted to the loop coordinates by minimizing the misalignment 
angles between the theoretical model and the observed loops, from which 
the photospheric force-free $\alpha$-parameter map $\alpha(x,y)$ and the 
photospheric current map $j_z(x,y)$ is derived (bottom).}
\end{figure}

\begin{figure}
\plotone{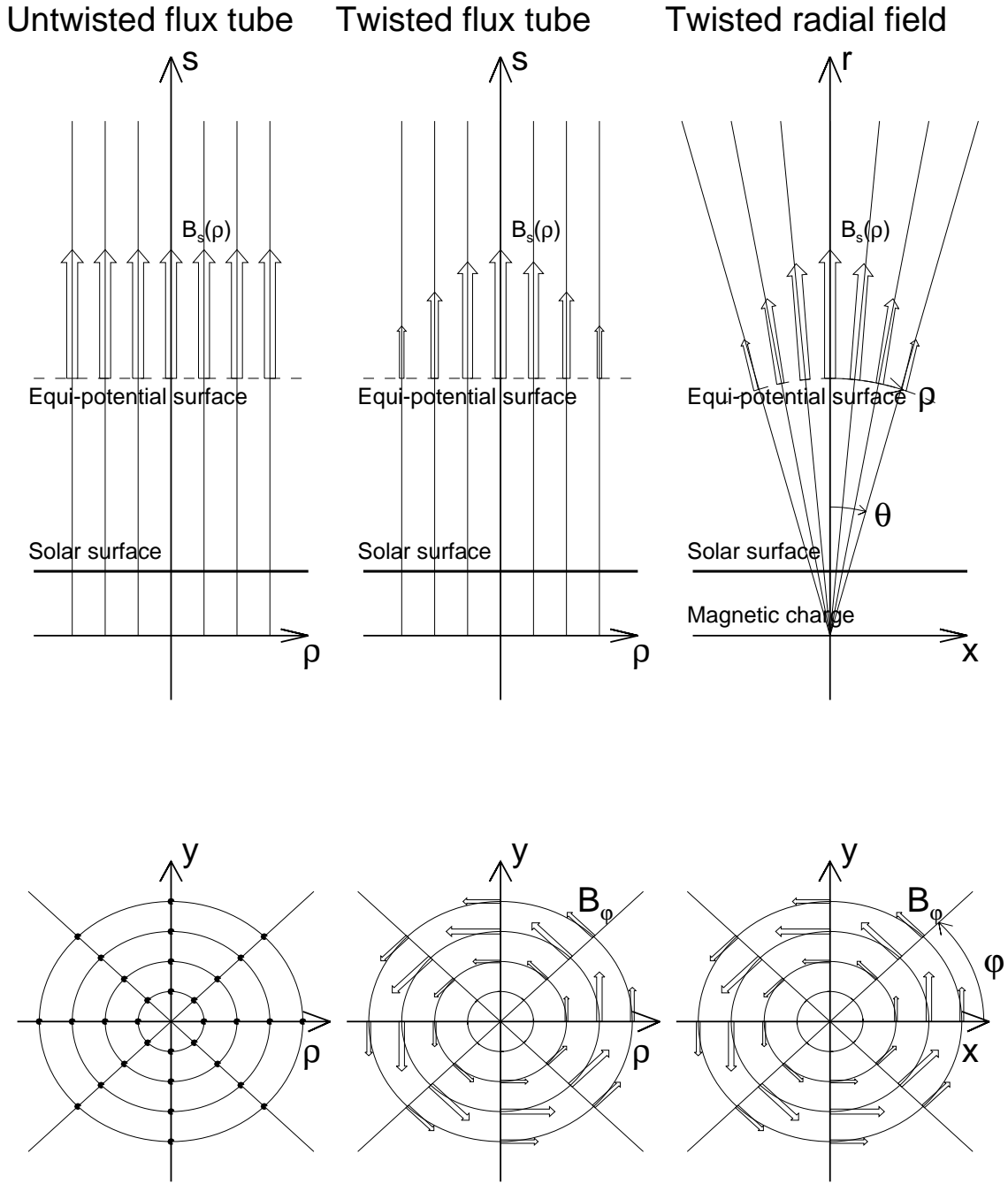}
\caption{The field line geometry is shown for an untwisted
cylindrical flux tube (left), a twisted cylindrical flux tube (middle),
and for a twisted radial field (right), from the side view in the
$xz$-plane (top) and from the top view in the $xy$-plane (bottom).
The top panels show the longitudinal magnetic field component $B_s(\rho)$
and the bottom panels show the azimuthal magnetic field component
$B_{\varphi}(\rho, \varphi)$.}
\end{figure}

\begin{figure}
\plotone{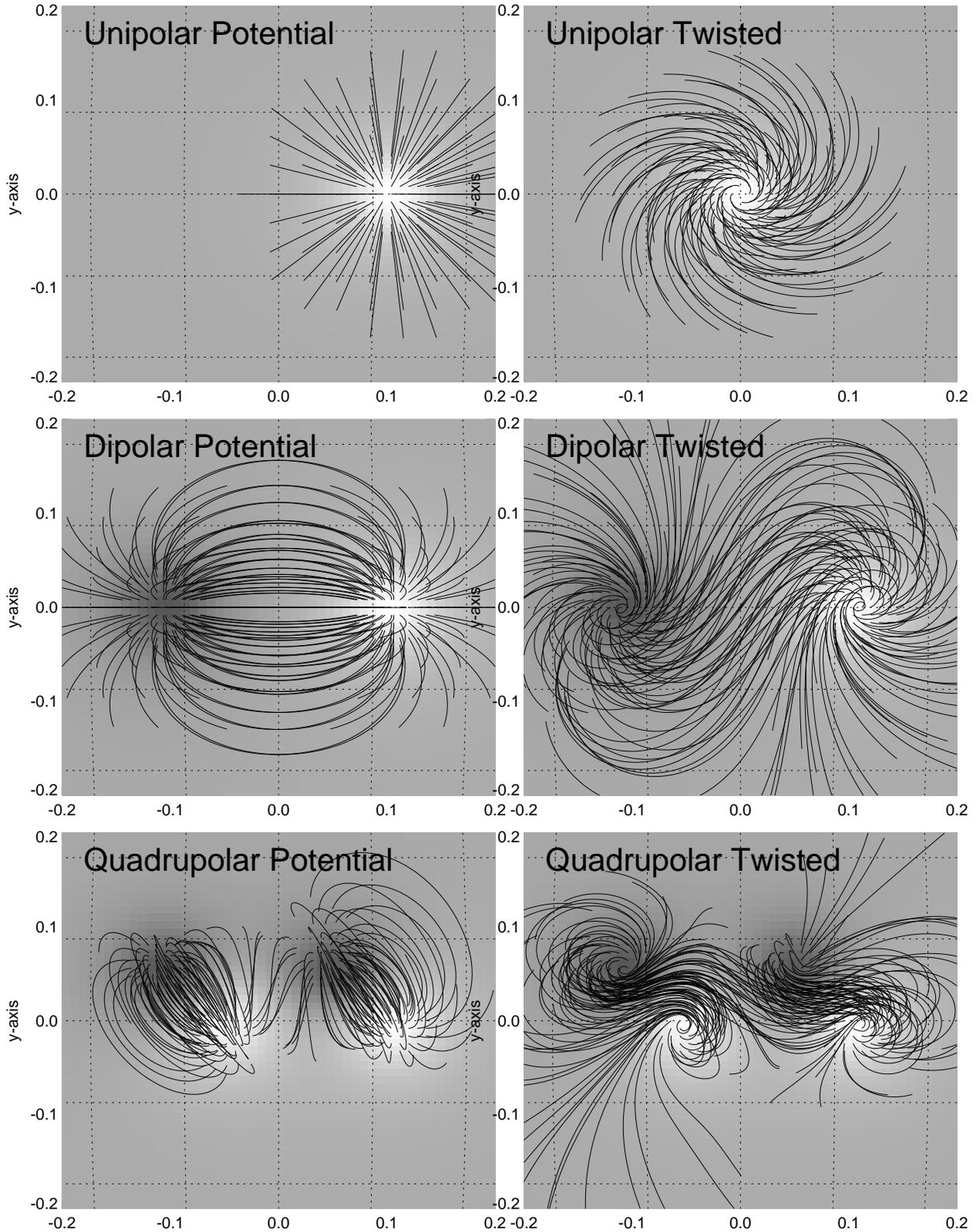}
\caption{Simulations of potential field($\alpha_j=0$; left side) 
and twisted non-potential field cases ($\alpha_j \ne 0$; right side), 
represented by line-of-sight magnetograms 
$B_z(x,y)$ (grey scale) and magnetic field lines $B(s)$ projected into 
the $x-y$ plane. Each set includes a unipolar magnetic charge (first row), 
a dipolar (second row), and a quadrupolar configuration (third row).}
\end{figure}

\begin{figure}
\plotone{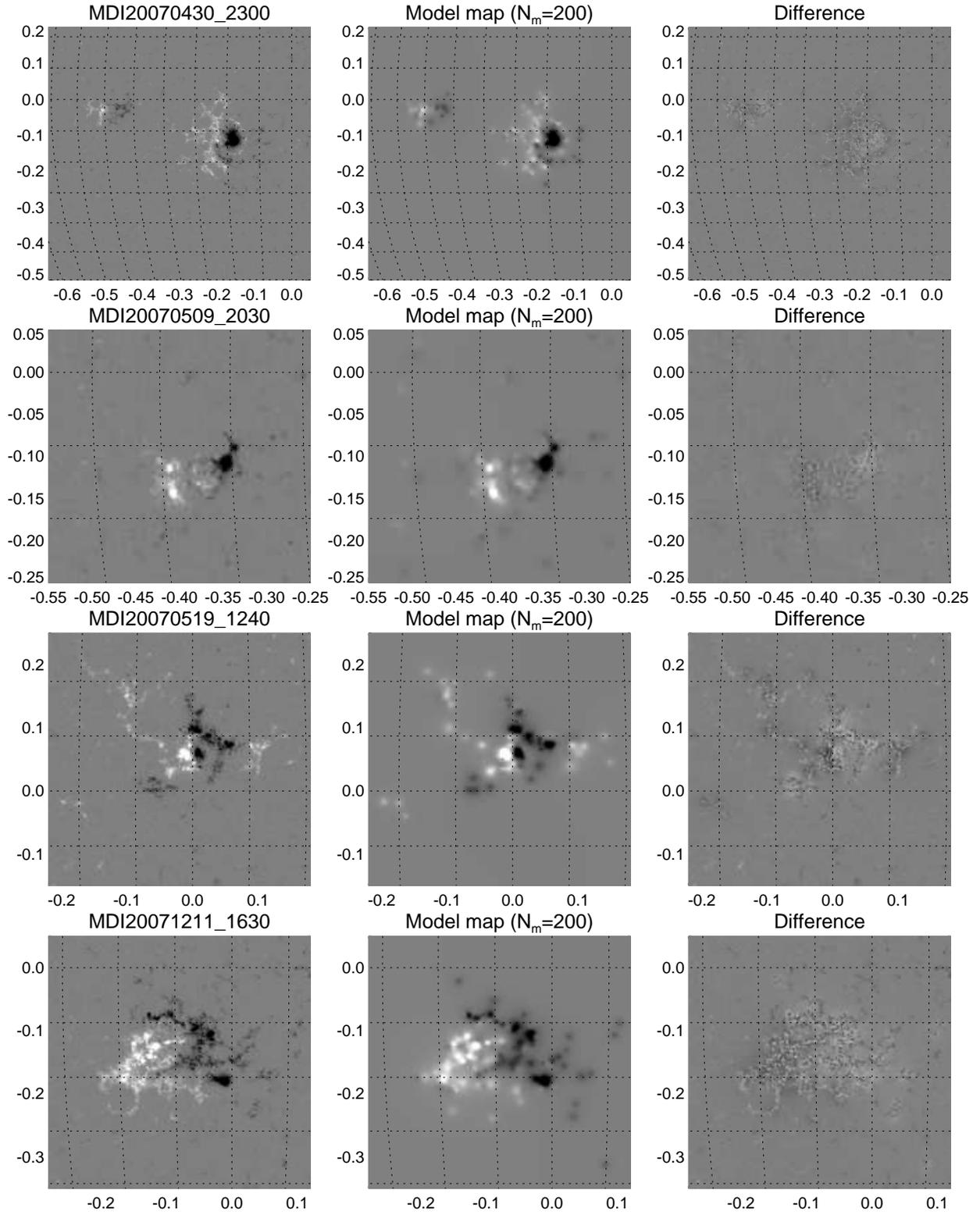}
\caption{Left column: photospheric magnetograms observed with {\sl SOHO}/MDI
for four observations. Middle column: the magnetic field is decomposed into
$N_m=200$ unipolar magnetic charges and the model map displays the
line-of-sight component of the magnetic field, $B_z$. Right column:
difference between the observed MDI magnetogram and the model.}
\end{figure}

\begin{figure}
\plotone{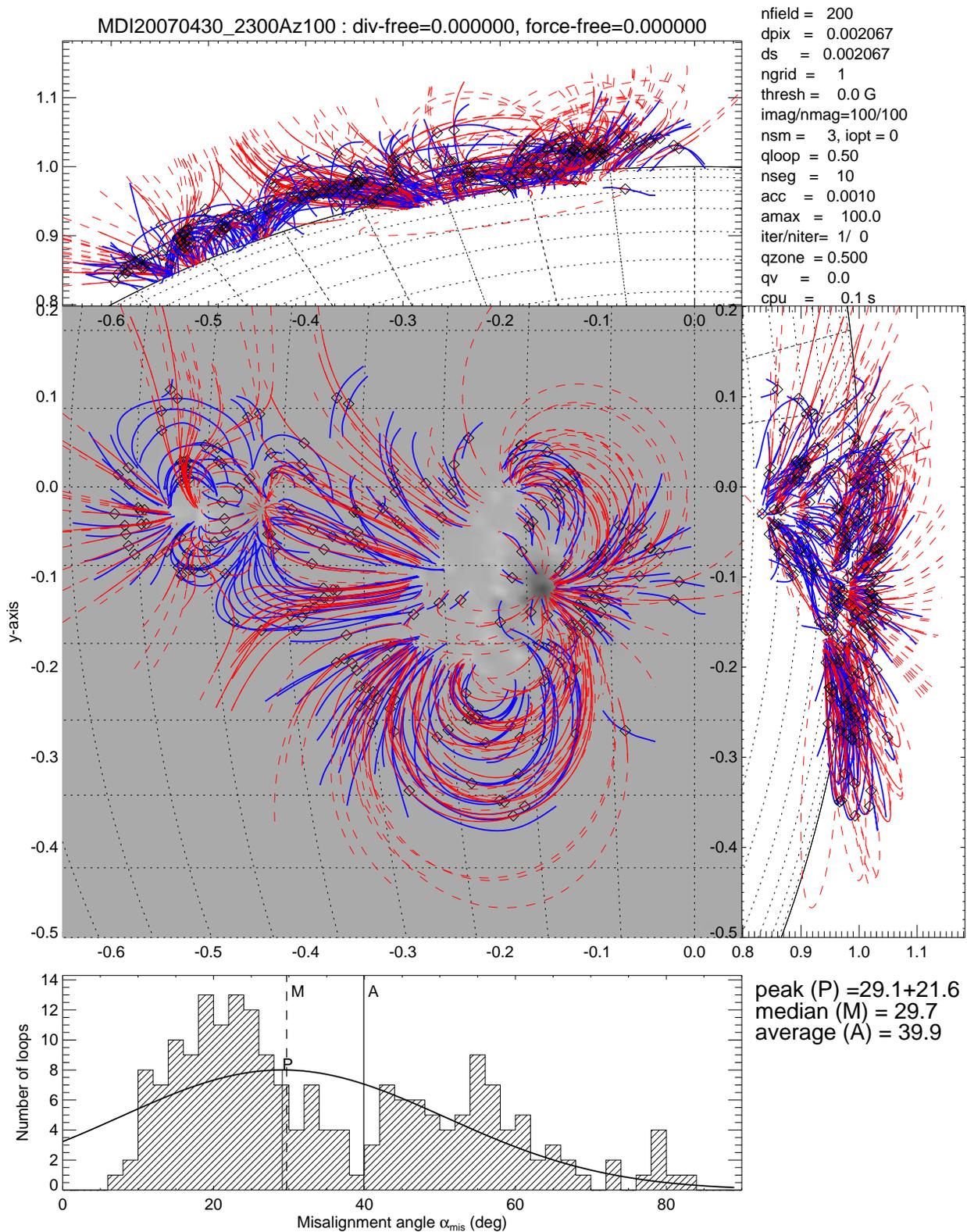}
\caption{A number of 200 stereoscopically triangulated loops observed
with STEREO (blue) are compared with the theoretical field lines of a 
potential field (red curves), defined by the potential fields of $N_m=100$ 
buried magnetic charges, decomposed from the MDI line-of-sight magnetogram.
The 3D views are shown in a top-down view in the $x-y$ plane (middle),
and in sideviews in $z-y$ plane (right) and $x-z$ plane (top). 
A histogram of the 3D misalignment angles is shown in the bottom panel, 
with the average (A), median (M), and peak value (P) of a fitted Gaussian
distribution. This active region NOAA 10953, labeled as $(A)$ in Table 1,
was observed on 2007 April 30, 23:00 UT, which is identical to the case
modeled in DeRosa et al. (2009).}
\end{figure}

\begin{figure}
\plotone{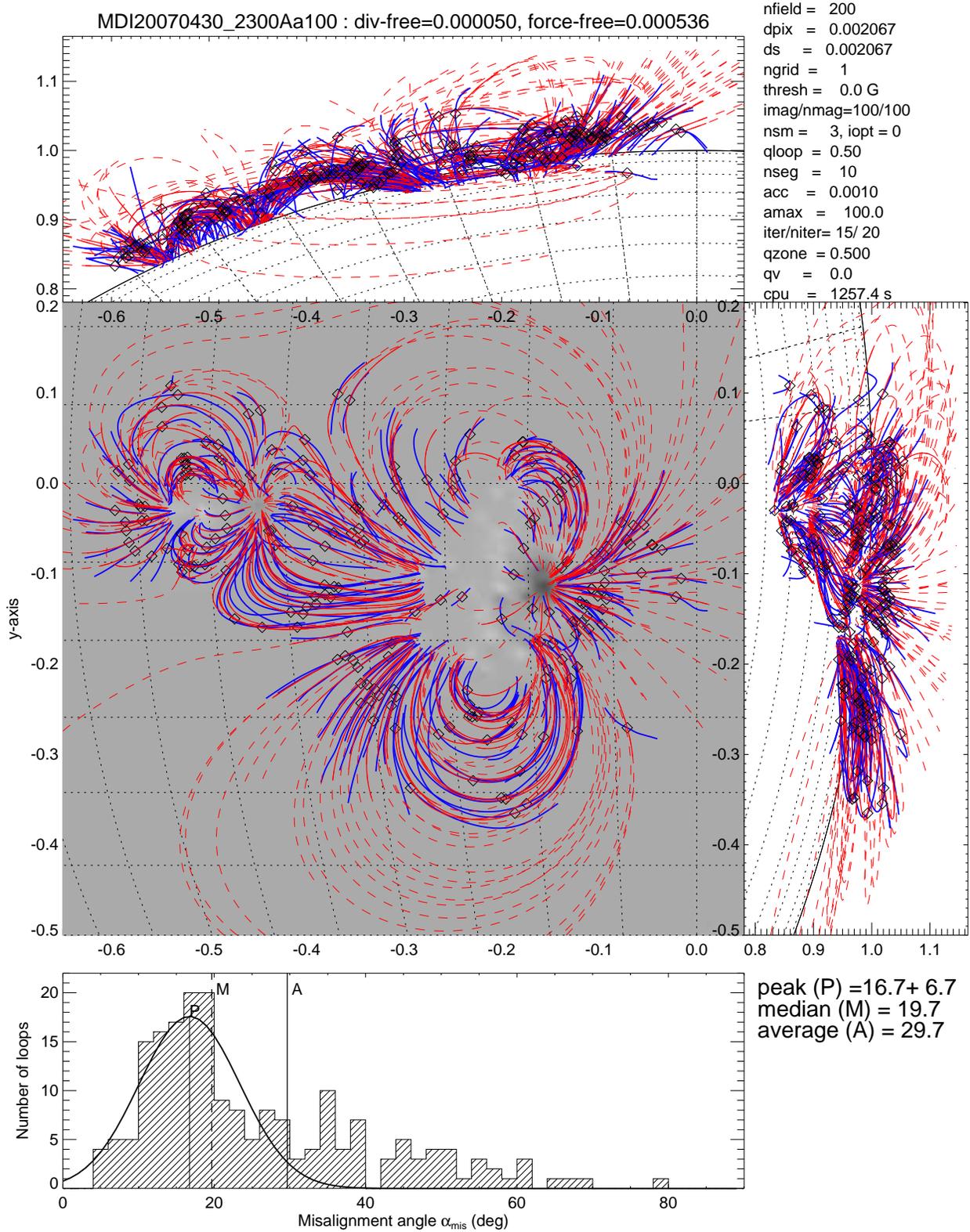}
\caption{Similar representation as in Fig.~5 for active region $A$
(2007 Apr 30), but modeled by forward-fitting our analytical approximation 
of a force-free field model. Note that the 3D misalignment angles for
this quasi-force-free field are significantly smaller than for the 
potential-field model shown in Fig.~5. 
See also movie A on forward-fitting of active region A.}
\end{figure}

\begin{figure}
\plotone{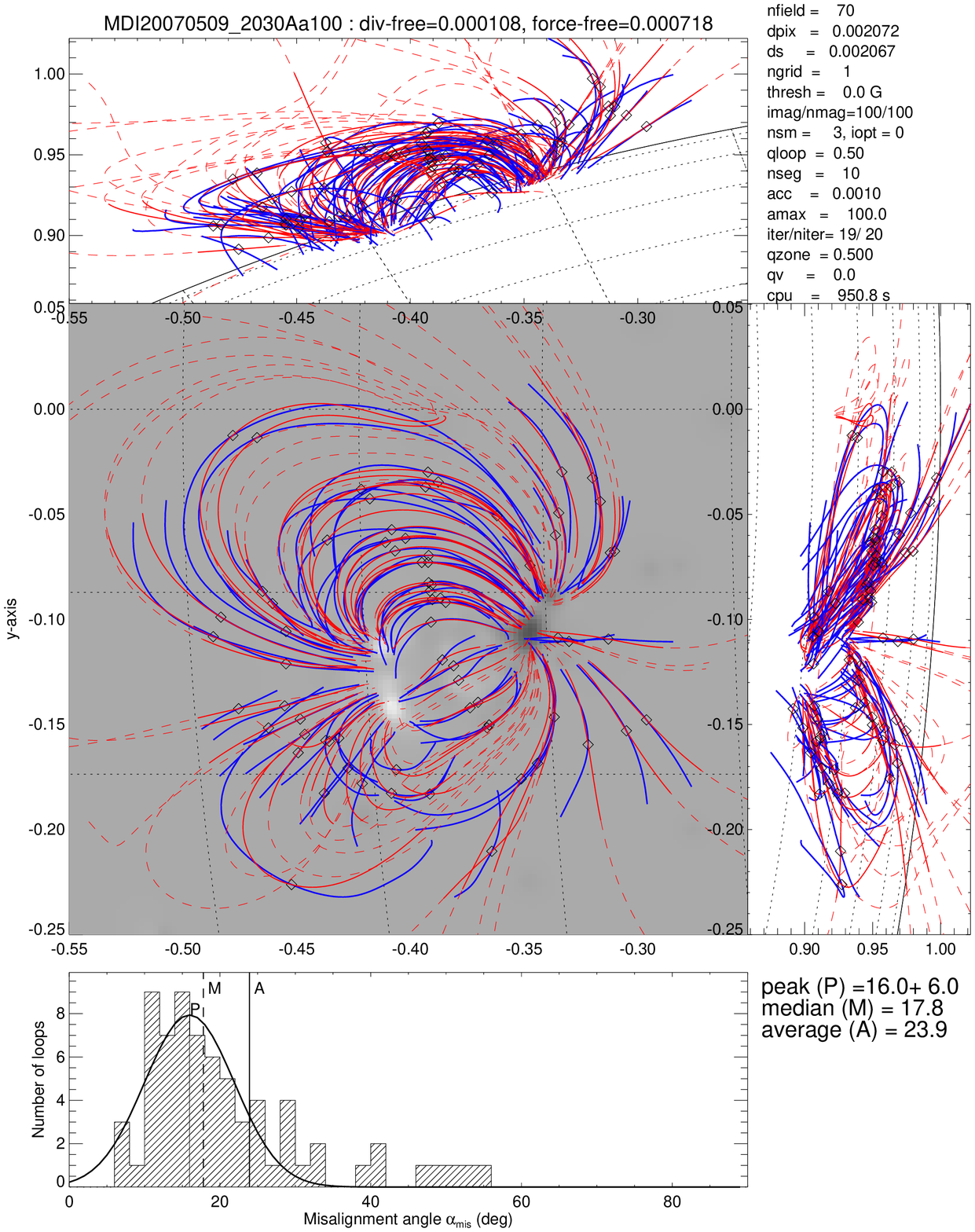}
\caption{Similar representation as in Fig.~6, for active region NOAA 10955
observed on 2007 May 9, 20:30 UT. 
See also movie B on forward-fitting of active region B.}
\end{figure}

\begin{figure}
\plotone{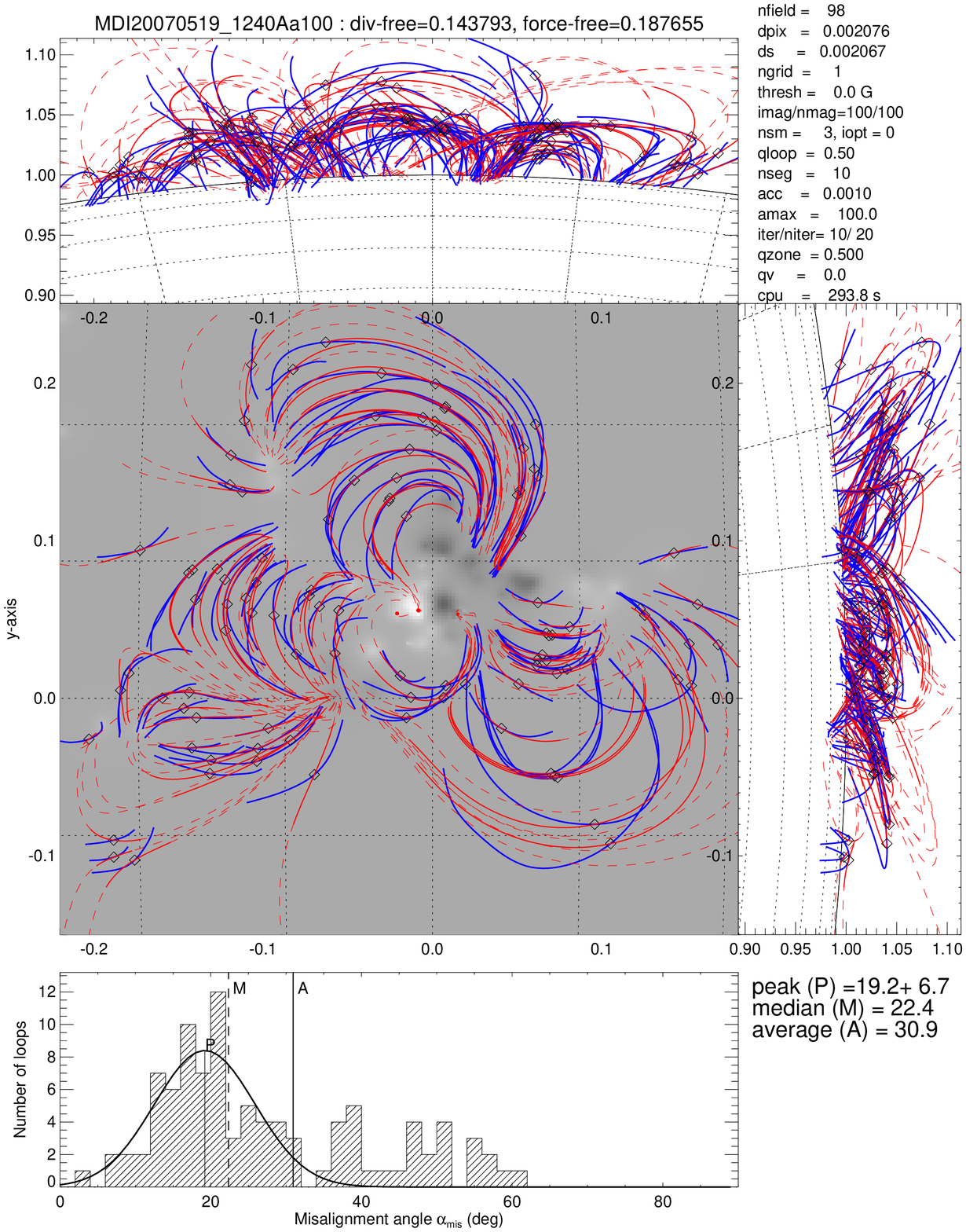}
\caption{Similar representation as in Fig.~6-7 for active region NOAA 10953
observed on 2007 May 19, 12:40 UT.
See also movie C on forward-fitting of active region C.}
\end{figure}

\begin{figure}
\plotone{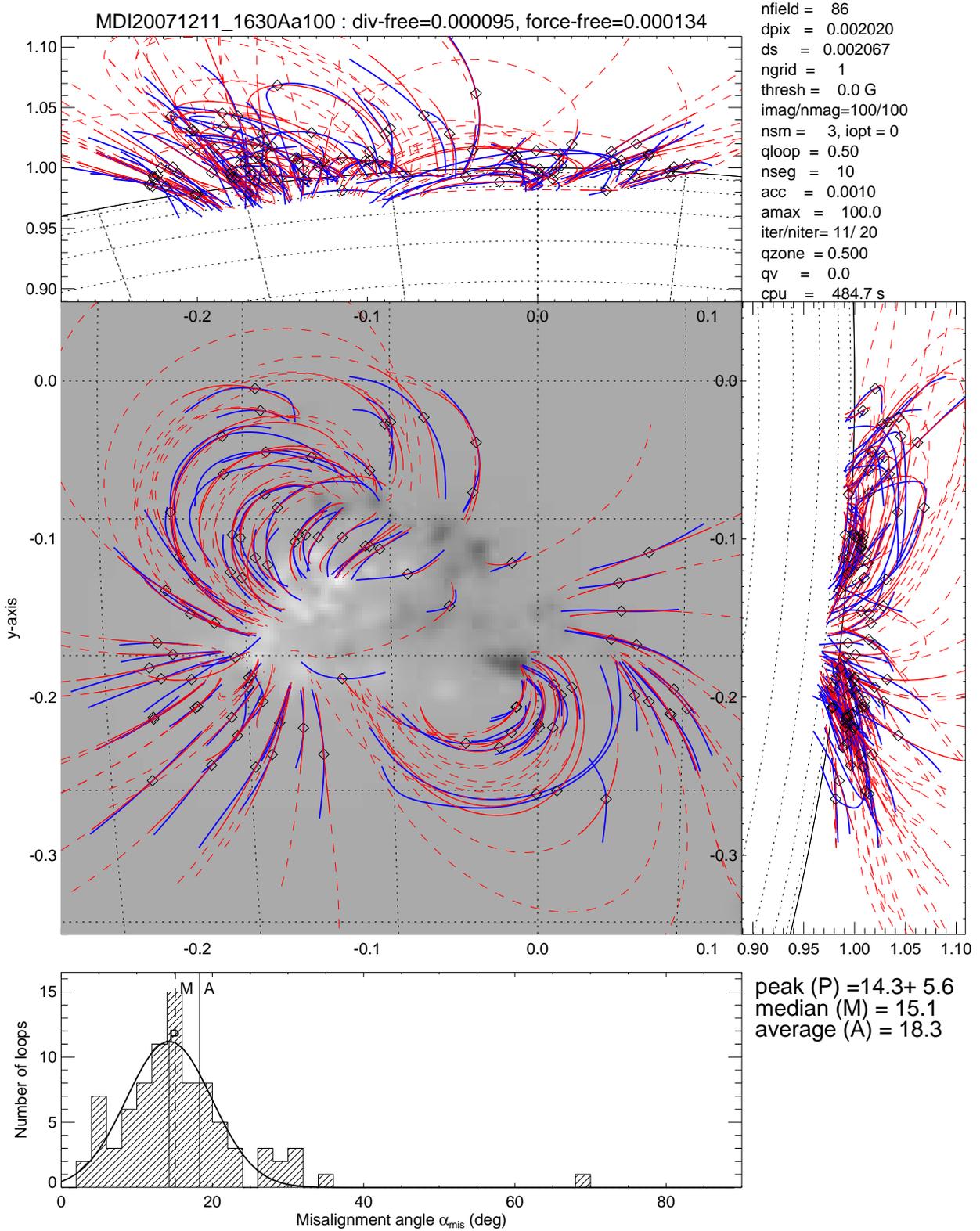}
\caption{Similar representation as in Fig.~6-7 for active region NOAA 10978
observed on 2007 Dec 11, 16:30 UT.
See also movie D on forward-fitting of active region D.}
\end{figure}

\begin{figure}
\plotone{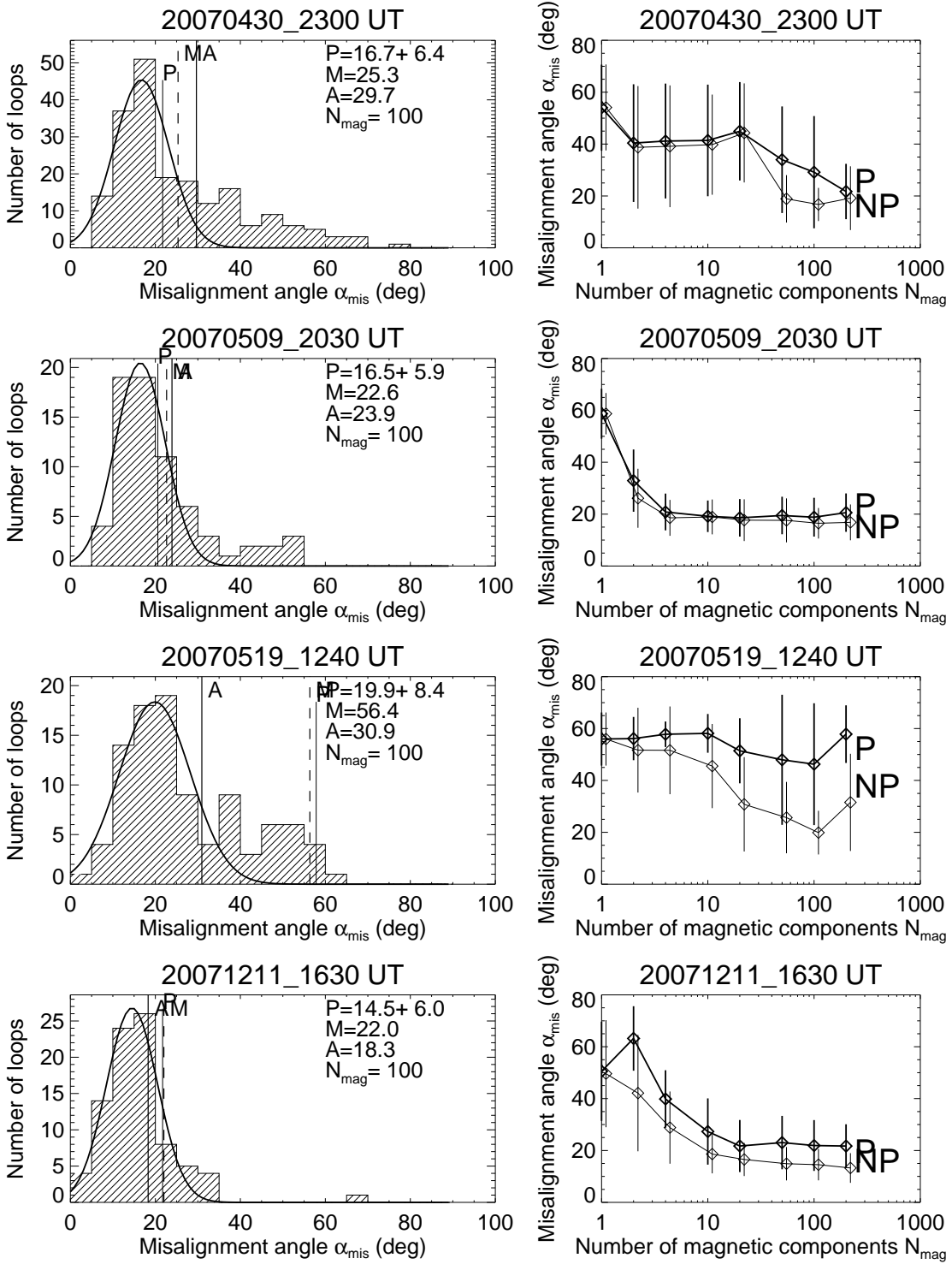}
\caption{The distribution of best-fit misalignment angles $\alpha_{mis}$
(between the observed loops and the best fit of the theoretical force-free
field model) is shown (left panels). The distributions are characterized by the
value of the peak (P) of a Gaussian fit with Gaussian width, by the
median (M), and the average (A). The dependence of the average misalignment
angle (A) as a function of the number of magnetic components $N_{m}$
is shown in the right panels, for the potential field model (P) and 
the non-potential (force-free field) model (NP).}
\end{figure}

\begin{figure}
\plotone{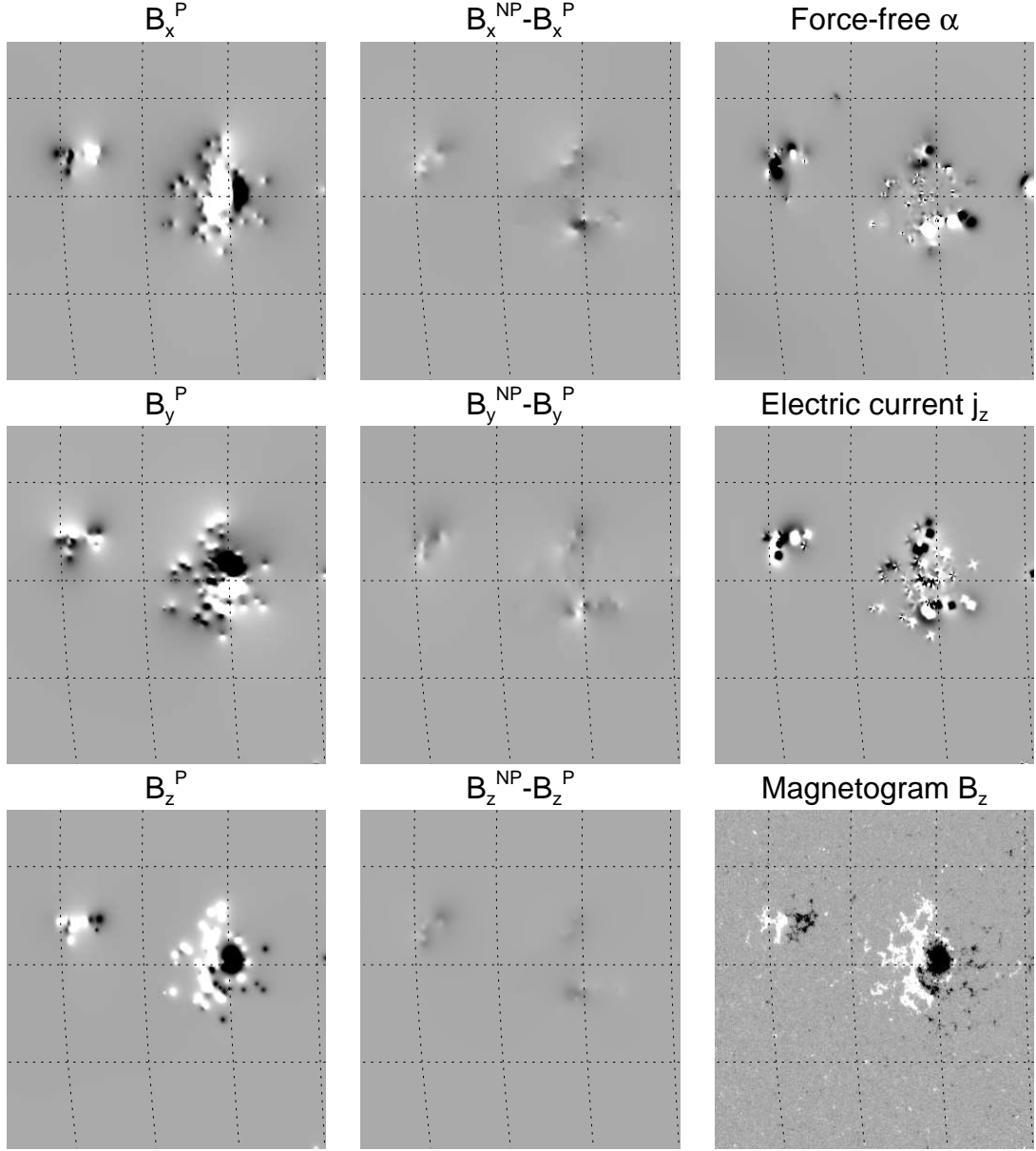}
\caption{The magnetic field component maps $B_x^P(x,y)$, $B_y^P(x,y)$,
and $B_z^P(x,y)$ are shown for the potential field model (left column),
difference maps of the non-potential field model to the potential field model,
$B_x^{NP}-B_x^{P}$, $B_y^{NP}-B_y^{P}$, and $B_z^{NP}-B_z^{P}$ (middle
column), the observed line-of-sight component $B_z^{obs}(x,y)$
(bottom right panel), and the derived nonlinear-force free $\alpha$-parameter
map $\alpha(x,y)$ (top right panel), and electric current $j_z(x,y)$ map (middle
right panel), for active region $A$ (2007 Apr 30). The grey scale is identical 
for the potential field and difference maps. The number of magnetic charge
components is $N_m=100$.}
\end{figure}

\begin{figure}
\plotone{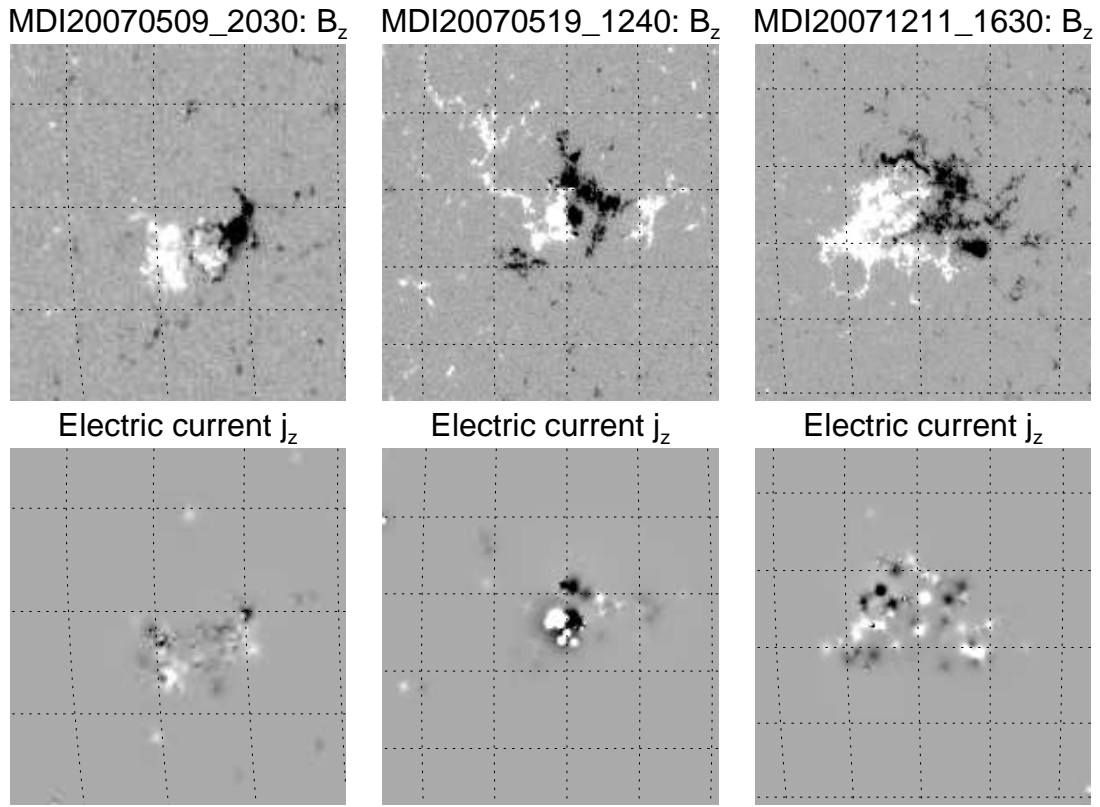}
\caption{Observed SOHO/MDI magnetogram $B_z(x,y)$ (top row) and 
electric current maps $j(x,y)$ obtained from the force-free field
forward-fitting, for active regions $B$ (2007 May 9), $C$ (2007 May 19), 
and $D$ (2007 Dec 11).}
\end{figure}

\begin{figure}
\plotone{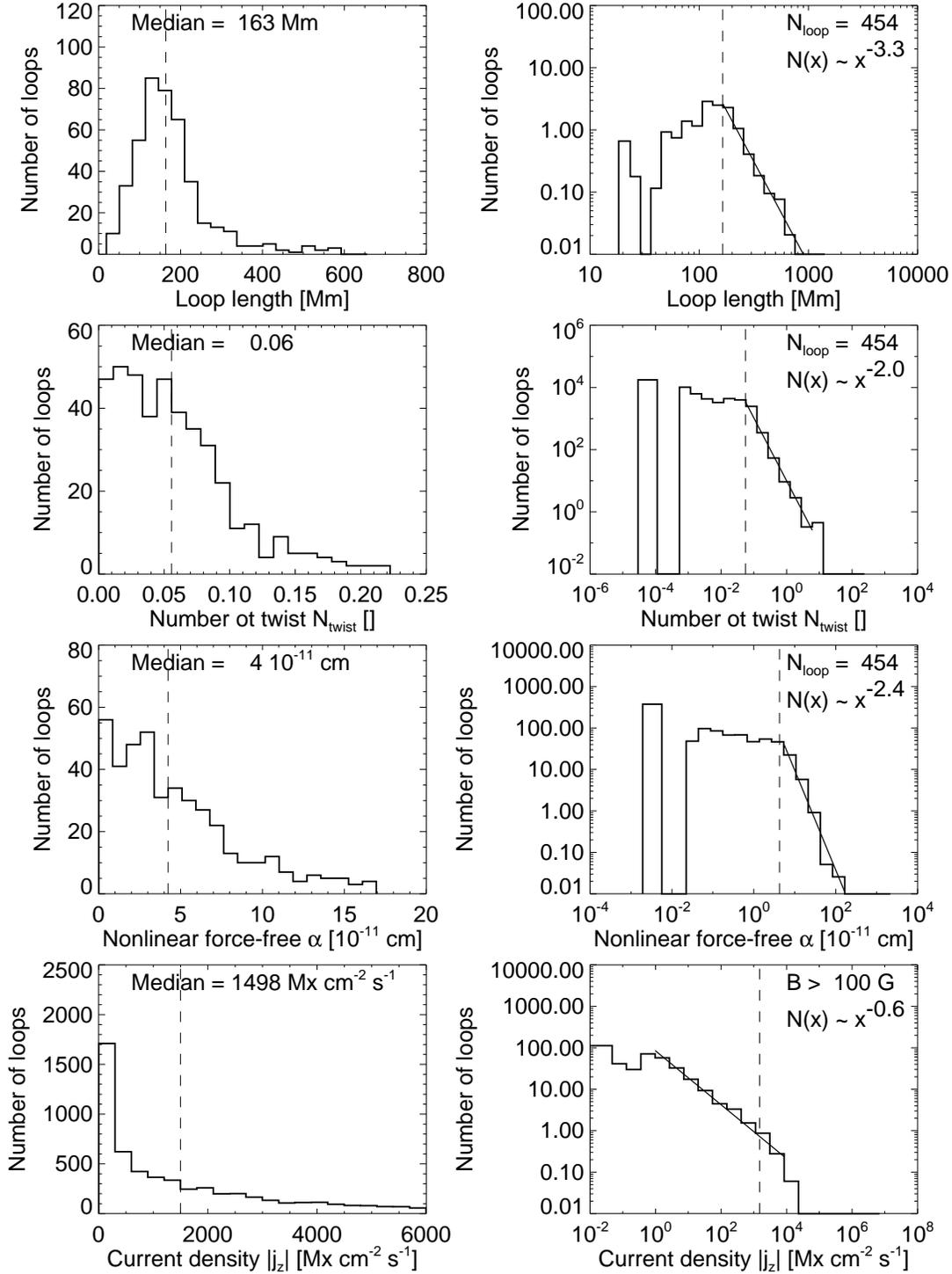}
\caption{Statistical distributions of loop-associated field line lengths $L$
(top row), the number of twist turns per loop, $N_{twist}$ (second row),
the force-free $\alpha$-parameter (third row), and the current density
$|j_z|$ (bottom row), in lin-lin (left column) and log-log histograms
(right column), with the median indicated (dashed vertical lines)
and powerlaw fits of the distributions.}
\end{figure}

\begin{figure}
\plotone{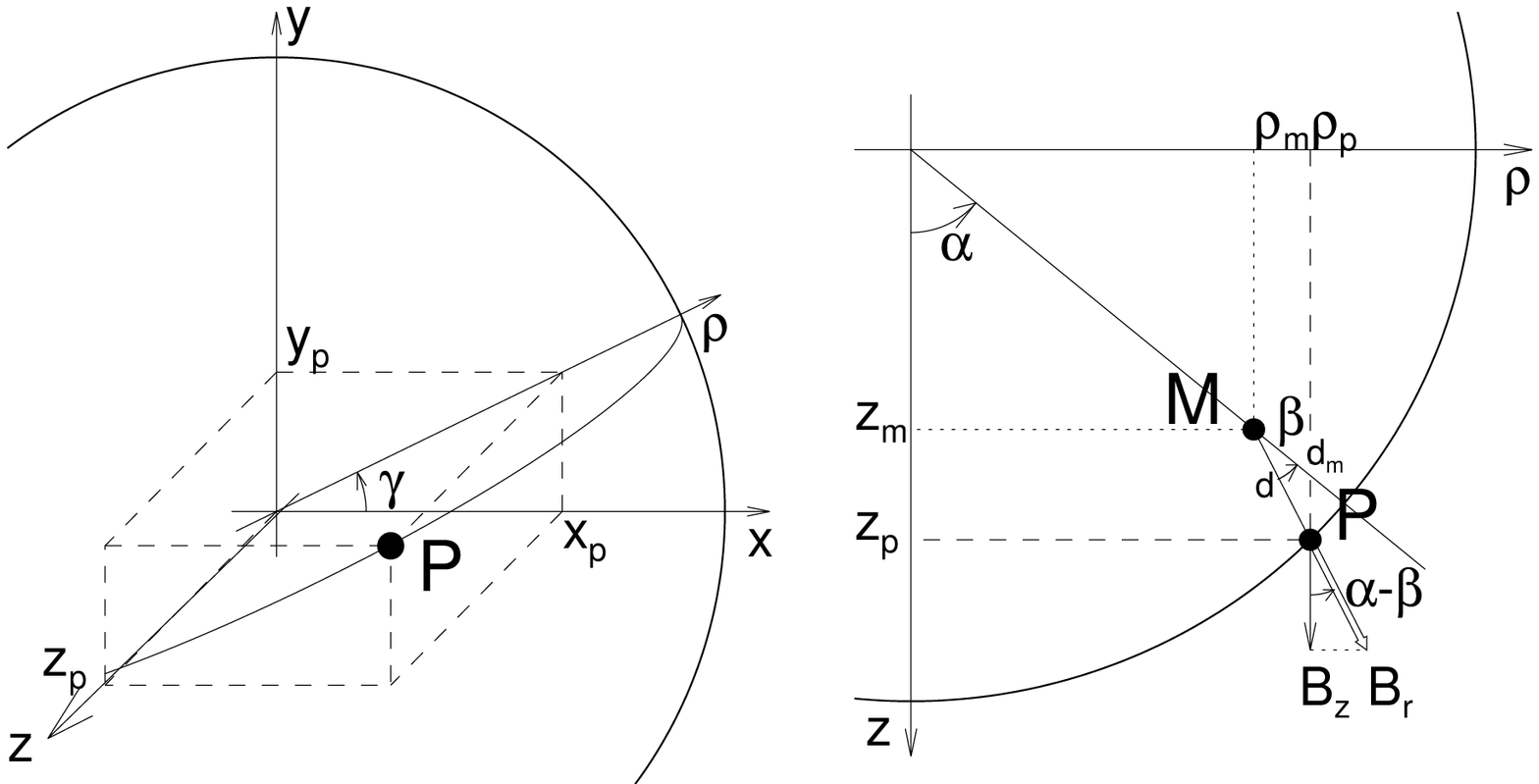}
\caption{3D geometry of a point source $P=(x_p, y_p, z_p)$ 
in a cartesian coordinate system is shown (left), with the $z$-axis aligned 
to the line-of-sight from Earth to Sun center. The plane through the 
line-of-sight axis and the point source $P$ has a position angle $\gamma$ 
in the plane-of-sky with respect to the $x$-axis and defines the direction
of the axis $\rho$.
The geometry of a line-of-sight magnetic field component $B_z$ is shown
in the $(z,\rho)$-plane on the right hand side. A magnetic point charge $M$
is buried at position $(z_m, \rho_m)$ and has an aspect angle $\alpha$
to the line-of-sight. The radial component $B_r$ is observed on the solar 
surface at location $P$ and has an inclination angle of $\beta$ to the local 
vertical above the magnetic point charge $M$. The line-of-sight component
$B_z$ of the magnetic field has an angle $(\alpha-\beta)$ to the radial
magnetic field component $B_r$. See details in Appendix A.}
\end{figure}


\begin{references}

\reference{}
	Altschuler, M.D. and Newkirk,G.Jr. 1969, \solphys 9, 131.
\reference{}
	Aschwanden, M.J., Kosugi, T., Hudson, H.S., Wills, M.J., and
        Schwartz, R.A. 1996, \apj 470, 1198.
\reference{}
	Aschwanden, M.J., Newmark, J.S., Delaboudiniere, J.P., 
	Neupert, W.M., Klimchuk, J.A., Gary, G.A., Portier-Fornazzi, F., 
	and Zucker, A., 1999, \apj 515, 842.
\reference{}
	Aschwanden, M.J. 2004, {\sl Physics of the Solar Corona. 
	An Introduction}, PRAXIS Publishing Co., Chichester UK, and 
	Springer, Berlin.
\reference{}
	Aschwanden,M.J., Lee,J.K., Gary,G.A., Smith,M., and Inhester,B.
        2008a, \solphys 248, 359.
\reference{}
	Aschwanden, M.J., Wuelser, J.P., Nitta, N.V., \& Lemen, J.R. 2008b,
        ApJ 679, 827 (Paper I).
\reference{}
	Aschwanden, M.J., Nitta, N.V., Wuelser, J.P., \& Lemen, J.R. 2008c,
        ApJ 680, 1477 (Paper II).
\reference{}
	Aschwanden, M.J., Wuelser, J.P., Nitta, N., Lemen, J., and Sandman, A.
        2009, \apj 695, 12 (Paper III).
\reference{}
	Aschwanden,M.J. 2010, \solphys 262, 399.
\reference{}
	Aschwanden, M.J. and Sandman, A.W. 2010, Astronomical J. 140, 723.
\reference{}
	Aschwanden, M.J. 2011, Living Reviews in Solar Physics 8, 5.
\reference{}
	Aschwanden,M.J. 2012, \solphys (in press), \\
        http://www.lmsal.com/$~{\ }$aschwand/eprints/2012$\_$fff1.pdf
\reference{}
	Aschwanden,M.J. and Malanushenko, A. 2012, \solphys (in press), \\
        http://www.lmsal.com/$~{\ }$aschwand/eprints/2012$\_$fff2.pdf.
\reference{}
	Berger, M.A. 1991, \aap 252, 369.
\reference{}
	Bobra, M.G., Van Ballegooijen, A.A., and DeLuca, E.E. 2008,
        \apj 672, 1209.
\reference{}
	Boyd, T.J.M. and Sanderson, J.J. 2003, {\sl The physics of plasmas},
        Cambridge University Press, Cambridge.
\reference{}
	Conlon, P.A. and Gallagher, P.T. 2010, \apj 715, 59.
\reference{}
	DeRosa, M.L., Schrijver, C.J., Barnes, G., Leka, K.D., Lites, B.W.,
        Aschwanden, M.J., Amari, T., Canou, A., McTiernan, J.M., Regnier, S.,
        Thalmann, J., Valori, G., Wheatland, M.S., Wiegelmann, T.,
        Cheung, M.C.M., Conlon, P.A., Fuhrmann, M., Inhester, B.,
        and Tadesse, T. 2009, \apj 696, 1780.
\reference{}
	Fan, Y. and Gibson, S.E. 2003, \apj 589, L105.
\reference{}
	Fan, Y. and Gibson, S.E. 2004, \apj 609, 1123.
\reference{}
	Feng, L., Wiegelmann, T., Inhester, B., Solanki, S., Gan, W.Q.,
        and Ruan, P. 2007a, \solphys 241, 235.
\reference{}
	Feng, L., Inhester, B., Solanki, S., Wiegelmann, T., Podlipnik, B.,
        Howard, R.S., Wuelser, J.P. 2007b, ApJ 671, L205
\reference{}
	Gold, T. and Hoyle, F. 1958, MNRAS 120, 89.
\reference{}
	Inhester, B., Feng, L., and Wiegelmann, T. 2008, \solphys 248, 379.
\reference{}
	Inverarity, G.W. and Priest, E.R. 1995, \aap 296, 395.
\reference{}
	Jing, J., Tan, C., Yuan, Y., Wang, B., Wiegelmann, T., Xu, Y., Wang, H.
        2010, \apj 713, 440.
\reference{}
	Kaiser, M.L., Kucera, T.A., Davila, J.M., St.Cyr, O.C., Guhathakurta,M.,
        and Christian, E. 2008, Space Science Reviews 136, 5.
\reference{}
	Kliem, B., Titov, V.S., and T\"or\"ok, T. 2004, \aap 413, L23.
\reference{}
	Leka, K.D., Canfield, R.C., and McClymont, A.N. 1996, \apj 462, 547.
\reference{}
	Longcope, D.W., Fisher, G.H., and Pevtsov, A.A. 1998, \apj 507, 417.
\reference{}
	Longcope, D.W. and Welsch, B.T. 2000, \apj 545, 1089.
\reference{}
	Longcope, D.W. 2005, Living Reviews in Solar Physics 2, 7.
\reference{}
	Low, B.C. and Lou, Y.Q. 1990, \apj {\bf 352}, 343.
\reference{}
	Luhmann, J.G., Gosling, J.T., Hoeksema, J.T., and Zhao, X.
        1998, \jgr 103(A4), 6585.
\reference{}
	Lundquist, L.L., Fisher, G.H., \& McTiernan, J.M. 2008a, ApJS 179, 509.
\reference{}
	Lundquist, L.L., Fisher, G.H., \& McTiernan, J.M. 2008b, ApJ 689, 1388.
\reference{}
	Malanushenko, A., Longcope, D.W., and McKenzi, D.E. 2009,
        \apj 707, 1044.
\reference{}
	Malanushenko, A., Longcope, D.W., and McKenzi, D.E. 2009,
	\apj 707, 1044.
\reference{}
	Malanushenko, A., Schrijver, C.J., DeRosa, M.L., Wheatland, M.S.,
	and Gilchrist, S.A. \apj, (in press).
\reference{}
	Mikic, Z., Schnack, D.D., and VanHoven,G. 1990, \apj 361, 690.
\reference{}
	Parker, E.N. 1988, \apj 330, 474.
\reference{}
	Petrie, G.J.D., Canou, A., and Amari, T. 2011, \solphys (in press).
\reference{}
	Pevtsov, A.A., Canfield, R.C., and McClymont, A.N. 1997,
        \apj 481, 973.
\reference{}
	Portier-Fozzani, F., Aschwanden, M.J., Demoulin, P., Neupert, W.,
        and EIT Team 2001, \solphys 203, 289.
\reference{}
	Press, W.H., Flannery, B.P., Teukolsky, S.A., and Vetterling, W.T. 1986,
        {\sl Numerical recipes. The Art of Scientific Computing},
        Cambridge University Press: New York.
\reference{}
	Priest, E.R. 1982, {\sl Solar Magnetohyrdodynamics},
        Geophysics and Astrophysics Monographs Volume 21,
        D. Reidel Publishing Company, Dordrecht.
\reference{}
	Priest, E.R., Hood, A.W., and Anzer, U. 1989, \apj 344, 1010.
\reference{}
	Priest, E.R., Van Ballegooijen, A.A., and MacKay, D.H. 1996,
        \apj 460, 530.
\reference{}
	Rosner, R., Tucker, W.H., and Vaiana, G.S. 1978, \apj 220, 643.
\reference{}
	Ruan, P., Wiegelmann, T., Inhester, B., Neukirch, T., Solanki, S.K.,
        and Feng, L. 2008, \aap 481, 827.
\reference{}
	Rust, D.M. and Kumar, A. 1996, \apj 464, L199.
\reference{}
	Sakurai, T. 1982, \solphys 76, 301.
\reference{}
	Sandman, A., Aschwanden, M.J., DeRosa, M., Wuelser, J.P. and
        Alexander, D. 2009, \solphys 259, 1.
\reference{}
	Sandman, A.W. and Aschwanden, M.J. 2011, \solphys 270, 503.
\reference{}
	Schrijver, C.J., Sandman, A.W., Aschwanden, M.J., and DeRosa, M.L.
        2004, \apj 615, 512.
\reference{}
	Schrijver, C.J., DeRosa, M., Metcalf, T.R., Liu, Y., McTiernan, J.,
        Regnier, S., Valori, G., Wheatland, M.S., and Wiegelmann, T.
        2006, \solphys 235, 161.
\reference{}
	Sturrock, P.A. 199i4, {\sl Plasma physics. An introduction to the
        theory of astrophysica, geophysical and laboratory plasmas},
        Cambridge University Press, Cambridge.
\reference{}
	Su, Y.N., Surges, V., vanBallegooijen, A.A., Deluca, E., and Golub,L.
        2011, \apj 734, 53.
\reference{}
	T\"or\"ok, T., and Kliem, B. 2003, \aap 406, 1043.
\reference{}
	Warren, H.P. and Winebarger, A.R. 2006, ApJ 645, 711.
\reference{}
	Warren, H.P. and Winebarger, A.R. 2007, ApJ 666, 1245.
\reference{}
	Warren, H.P., Winebarger, A.R., and Brooks, D.H. 2010, \apj 711, 228.
\reference{}
	Wiegelmann, T. and Neukirch, T. 2002, \solphys 208, 233.
\reference{}
	Wiegelmann, T. and Inhester, B. 2003, \solphys 214, 287.
\reference{}
	Wiegelmann, T., Lagg, A., Solanki, S.K., Inhester, B., and Woch, J.
        2005 \aap 433, 701.
\reference{}
	Wiegelmann, T. and Inhester, B. 2006, \solphys 236, 25.
\reference{}
	Wiegelmann, T., Inhester, B., and Feng, L. 2009, Annales Geophysicae
        27/7, 2925.
\reference{}
	Withbroe, G.L. and Noyes, R.W. 1977, Ann. Rev. Astron. Astrophys.
        15, 363.

\end{references}
\end{document}